\documentclass[onecolumn]{aastex631}

\usepackage{gensymb}

\def\amin{\ifmmode^{\prime}\else$^{\prime}$\fi}
\def\asec{\ifmmode^{\prime\prime}\else$^{\prime\prime}$\fi}

\newcommand{\psr}{PSR~J2229$+$6114}

\newcommand{\nustar}{\textit{NuSTAR}}
\newcommand{\xmm}{{\it XMM-Newton}}

\newcommand{\chandra}{{\it Chandra}}
\newcommand{\fermi}{{\it Fermi-}LAT}
\newcommand{\veritas}{VERITAS} 

\newcommand{\suzaku}{\textit{Suzaku}}

\newcommand{\fluxcgs}{ergs~s$^{-1}$~cm$^{-2}$}
\newcommand{\lumcgs}{ergs~s$^{-1}$}

\newcommand{\src}{G106.6+2.9}

\def\s{\phantom}

\def\amin{\ifmmode^{\prime}\else$^{\prime}$\fi}
\def\asec{\ifmmode^{\prime\prime}\else$^{\prime\prime}$\fi}

\def\simgt{\lower.5ex\hbox{$\; \buildrel > \over \sim \;$}}
\def\simlt{\lower.5ex\hbox{$\; \buildrel < \over \sim \;$}}

\def\asec{\ifmmode^{\prime\prime}\else$^{\prime\prime}$\fi}

\newcommand{\DM}{cm$^{-3}$\,pc}

\shorttitle{Boomerang}
\shortauthors{Pope et al.}
\graphicspath{{./}{figures/}}

\begin{document}

\title{A multi-wavelength investigation of \psr\ and its pulsar wind nebula in the radio, X-ray, and gamma-ray bands}

\author{I.~Pope}\thanks{I.~Pope, isaac.pope@columbia.edu}\affiliation{Columbia Astrophysics Laboratory, 550 West 120th Street, New York, NY 10027, USA}
\author{K.~Mori}\affiliation{Columbia Astrophysics Laboratory, 550 West 120th Street, New York, NY 10027, USA}
\author{M.~Abdelmaguid}\affiliation{NYU Abu Dhabi, PO Box 129188, Abu Dhabi, United Arab Emirates}
\author{J.~D.~Gelfand}\affiliation{NYU Abu Dhabi, PO Box 129188, Abu Dhabi, United Arab Emirates}
\author{S.~P.~Reynolds}\affiliation{Physics Department, NC State University, Raleigh, NC 27695, USA}
\author{S.~Safi-Harb}\affiliation{Department of Physics and Astronomy, University of Manitoba, Winnipeg, MB R3T 2N2, Canada}
\author{C.~J.~Hailey}\affiliation{Columbia Astrophysics Laboratory, 550 West 120th Street, New York, NY 10027, USA}
\author{H.~An}\affiliation{Chungbuk National University, Chungdae-ro 1, Seowon-Gu, Cheongju, Chungbuk, 28644 South Korea}
\collaboration{8}{(\nustar\ Collaboration)}

\author{P.~Bangale}\affiliation{Department of Physics and Astronomy and the Bartol Research Institute, University of Delaware, Newark, DE 19716, USA}
\author{P.~Batista}\affiliation{DESY, Platanenallee 6, 15738 Zeuthen, Germany}
\author{W.~Benbow}\affiliation{Center for Astrophysics $|$ Harvard \& Smithsonian, Cambridge, MA 02138, USA}
\author{J.~H.~Buckley}\affiliation{Department of Physics, Washington University, St. Louis, MO 63130, USA}
\author{M.~Capasso}\thanks{M.~Capasso, capasso@nevis.columbia.edu}\affiliation{Department of Physics and Astronomy, Barnard College, Columbia University, NY 10027, USA}
\author{J.~L.~Christiansen}\affiliation{Physics Department, California Polytechnic State University, San Luis Obispo, CA 94307, USA}
\author{A.~J.~Chromey}\affiliation{Center for Astrophysics $|$ Harvard \& Smithsonian, Cambridge, MA 02138, USA}
\author{A.~Falcone}\affiliation{Department of Astronomy and Astrophysics, 525 Davey Lab, Pennsylvania State University, University Park, PA 16802, USA}
\author{Q.~Feng}\affiliation{Center for Astrophysics $|$ Harvard \& Smithsonian, Cambridge, MA 02138, USA}
\author{J.~P.~Finley}\affiliation{Department of Physics and Astronomy, Purdue University, West Lafayette, IN 47907, USA}
\author{G.~M~Foote}\affiliation{Department of Physics and Astronomy and the Bartol Research Institute, University of Delaware, Newark, DE 19716, USA}
\author{G.~Gallagher}\affiliation{Department of Physics and Astronomy, Ball State University, Muncie, IN 47306, USA}
\author{W.~F~Hanlon}\affiliation{Center for Astrophysics $|$ Harvard \& Smithsonian, Cambridge, MA 02138, USA}
\author{D.~Hanna}\affiliation{Physics Department, McGill University, Montreal, QC H3A 2T8, Canada}
\author{O.~Hervet}\affiliation{Santa Cruz Institute for Particle Physics and Department of Physics, University of California, Santa Cruz, CA 95064, USA}
\author{J.~Holder}\affiliation{Department of Physics and Astronomy and the Bartol Research Institute, University of Delaware, Newark, DE 19716, USA}
\author{T.~B.~Humensky}\affiliation{Department of Physics, University of Maryland, College Park, MD 20742-4111, USA}\affiliation{NASA GSFC, Greenbelt, MD 20771, USA}
\author{W.~Jin}\affiliation{Department of Physics and Astronomy, University of Alabama, Tuscaloosa, AL 35487, USA}
\author{P.~Kaaret}\affiliation{Department of Physics and Astronomy, University of Iowa, Van Allen Hall, Iowa City, IA 52242, USA}
\author{M.~Kertzman}\affiliation{Department of Physics and Astronomy, DePauw University, Greencastle, IN 46135-0037, USA}
\author{D.~Kieda}\affiliation{Department of Physics and Astronomy, University of Utah, Salt Lake City, UT 84112, USA}
\author{T.~K.~Kleiner}\affiliation{DESY, Platanenallee 6, 15738 Zeuthen, Germany}
\author{N.~Korzoun}\affiliation{Department of Physics and Astronomy and the Bartol Research Institute, University of Delaware, Newark, DE 19716, USA}
\author{F.~Krennrich}\affiliation{Department of Physics and Astronomy, Iowa State University, Ames, IA 50011, USA}
\author{S.~Kumar}\affiliation{Department of Physics, University of Maryland, College Park, MD 20742-4111, USA}
\author{M.~J.~Lang}\affiliation{School of Natural Sciences, University of Galway, University Road, Galway, H91 TK33, Ireland}
\author{G.~Maier}\affiliation{DESY, Platanenallee 6, 15738 Zeuthen, Germany}
\author{C.~E~McGrath}\affiliation{School of Physics, University College Dublin, Belfield, Dublin 4, Ireland}
\author{C.~L.~Mooney}\affiliation{Department of Physics and Astronomy and the Bartol Research Institute, University of Delaware, Newark, DE 19716, USA}
\author{P.~Moriarty}\affiliation{School of Natural Sciences, University of Galway, University Road, Galway, H91 TK33, Ireland}
\author{R.~Mukherjee}\affiliation{Department of Physics and Astronomy, Barnard College, Columbia University, NY 10027, USA}
\author{S.~O'Brien}\affiliation{Physics Department, McGill University, Montreal, QC H3A 2T8, Canada}
\author{R.~A.~Ong}\affiliation{Department of Physics and Astronomy, University of California, Los Angeles, CA 90095, USA}
\author{N.~Park}\thanks{N.~Park, nahee.park@queensu.ca}\affiliation{Department of Physics, Engineering Physics \& Astronomy, Queen's University, Kingston, ON K7L 3N6, Canada}
\author{S.~R.~Patel}\affiliation{DESY, Platanenallee 6, 15738 Zeuthen, Germany}
\author{K.~Pfrang}\affiliation{DESY, Platanenallee 6, 15738 Zeuthen, Germany}
\author{M.~Pohl}\affiliation{Institute of Physics and Astronomy, University of Potsdam, 14476 Potsdam-Golm, Germany and DESY, Platanenallee 6, 15738 Zeuthen, Germany}
\author{E.~Pueschel}\affiliation{DESY, Platanenallee 6, 15738 Zeuthen, Germany}
\author{J.~Quinn}\affiliation{School of Physics, University College Dublin, Belfield, Dublin 4, Ireland}
\author{K.~Ragan}\affiliation{Physics Department, McGill University, Montreal, QC H3A 2T8, Canada}
\author{P.~T.~Reynolds}\affiliation{Department of Physical Sciences, Munster Technological University, Bishopstown, Cork, T12 P928, Ireland}
\author{E.~Roache}\affiliation{Center for Astrophysics $|$ Harvard \& Smithsonian, Cambridge, MA 02138, USA}
\author{I.~Sadeh}\affiliation{DESY, Platanenallee 6, 15738 Zeuthen, Germany}
\author{L.~Saha}\affiliation{Center for Astrophysics $|$ Harvard \& Smithsonian, Cambridge, MA 02138, USA}
\author{G.~H.~Sembroski}\affiliation{Department of Physics and Astronomy, Purdue University, West Lafayette, IN 47907, USA}
\author{D.~Tak}\affiliation{DESY, Platanenallee 6, 15738 Zeuthen, Germany}
\author{J.~V.~Tucci}\affiliation{Department of Physics, Indiana University-Purdue University Indianapolis, Indianapolis, IN 46202, USA}
\author{A.~Weinstein}\affiliation{Department of Physics and Astronomy, Iowa State University, Ames, IA 50011, USA}
\author{D.~A.~Williams}\affiliation{Santa Cruz Institute for Particle Physics and Department of Physics, University of California, Santa Cruz, CA 95064, USA}
\author{J.~Woo}\thanks{J.~Woo, jw3855@columbia.edu}\affiliation{Columbia Astrophysics Laboratory, 550 West 120th Street, NY 10027, USA}
\collaboration{50}{(\veritas\ Collaboration)}





\begin{abstract}

G106.3$+$2.7, commonly considered a composite supernova remnant (SNR), is characterized by a boomerang-shaped pulsar wind nebula (PWN) and two distinct (“head” \& “tail”) regions in the radio band. A discovery of very-high-energy (VHE) gamma-ray emission ($E_\gamma > 100$ GeV) followed by the recent detection of ultra-high-energy (UHE) gamma-ray emission ($E_\gamma > 100$ TeV) from the tail region suggests that G106.3$+$2.7 is a PeVatron candidate. We present a comprehensive multi-wavelength study of the Boomerang PWN (100\asec\ around PSR J2229+6114) using archival radio and \chandra\ data obtained from two decades ago, a new \nustar\ X-ray observation from 2020, and upper limits on gamma-ray fluxes obtained by \fermi\ and VERITAS observatories. The \nustar\ observation allowed us to detect a 51.67 ms spin period from the pulsar PSR J2229+6114 and the PWN emission characterized by a power-law model with $\Gamma = 1.52\pm0.06$ up to 20 keV. Contrary to the previous radio study by \citet{Kothes2006}, we prefer a much lower PWN B-field ($B\sim3$ $\mu$G) and larger distance ($d \sim 8$ kpc) based on (1) the non-varying X-ray flux over the last two decades, (2) the energy-dependent X-ray PWN size resulting from synchrotron burn-off and (3) the multi-wavelength spectral energy distribution (SED) data. Our SED model suggests that the PWN is currently re-expanding after being compressed by the SNR reverse shock $\sim 1000$ years ago. In this case, the head region should be formed by GeV--TeV electrons injected earlier by the pulsar propagating into the low density environment.
\end{abstract}


\section{Introduction}
\label{sec:intro}

Pulsar wind nebulae \citep[PWNe; ][]{Cholis2018} are believed to generate a majority of the energetic leptons in our galaxy. The pulsar's rotating magnetic fields produce a wind of highly relativistic particles that expand out into the shell of the supernova remnant (SNR).  
High-energy observations of dozens of PWNe detected synchrotron and inverse-Compton upscattering (ICS) of cosmic microwave background photons, ambient infrared (IR) or optical stellar radiation in the X-ray and TeV bands, respectively, suggesting that non-thermal particles are accelerated to TeV or even PeV energies within the PWNe \citep{Arons2012}. 
The PWN evolution is characterized by three stages: (1) young, termination-shock driven wind nebulae, (2) middle-aged PWNe interacting with their host SNRs, and (3) relic PWNe \citep{Gaensler&Slane2006, Giacinti2020}. Relativistic winds from the pulsar injected into the SNR center are abruptly decelerated in an inward-facing termination shock, at which particles are accelerated to TeV energies; the post-shock flow further decelerates until it reaches pressure equilibrium with the SNR interior. The bubble of shocked pulsar wind is the observed PWN, which continues to expand until the deceleration of the outer SNR blast wave sends a reverse shock back toward the center, compressing and re-brightening the PWN, at an age of order 1--10 kyr. The PWN
continues to interact with the SNR interior until either the SNR dissipates or the pulsar, if born with a substantial kick velocity, escapes the SNR shell, continuing to inflate a PWN. In either case the PWN now interacts directly with the interstellar medium (ISM) \citep[a "relic PWN"; ][]{Cholis2018}, often in the shape of a bow-shock nebula. These middle-aged PWNe manifest a vast diversity of highly anisotropic non-thermal emission in multi-wavelength bands. The composite system is formed by its relic PWN interacting with the ambient medium and SNR reverse shock, exhibiting peculiar radio and X-ray morphology (often with nicknames such as Rabbit and Snail). Composite SNRs are of particular interest because they manifest on-going PWN-SNR interaction sites and possibly accelerate particles to TeV-PeV energies \citep{Ohira2018}. Some of the middle-aged PWNe are associated with the PeVatron candidates detected by HAWC and LHAASO above $E_{\gamma} \sim100$ TeV  \citep{Abeysekara2020, Cao2021}. Eventually, after $\tau \sim100$ kyr, electrons and positrons escape from relic PWNe and form extended TeV halos, as revealed around the Geminga and Monogem pulsars \citep{Abeysekara2017}. TeV halos are a new class of gamma-ray sources which are suggested to be the primary source of the positron excess observed at Earth \citep{LC2022}. How and when a PWN evolves through these transitions depends on the progenitor star's characteristic properties and environment within the ISM. Hence, multi-wavelength observations of PWNe in different evolution stages and environments are essential for understanding how particles are injected from the pulsar, diffuse out while cooling, and interact with the ambient gas and their host SNRs.       

\begin{figure}[t!]
\begin{center} 
  \includegraphics[width=0.75\textwidth]{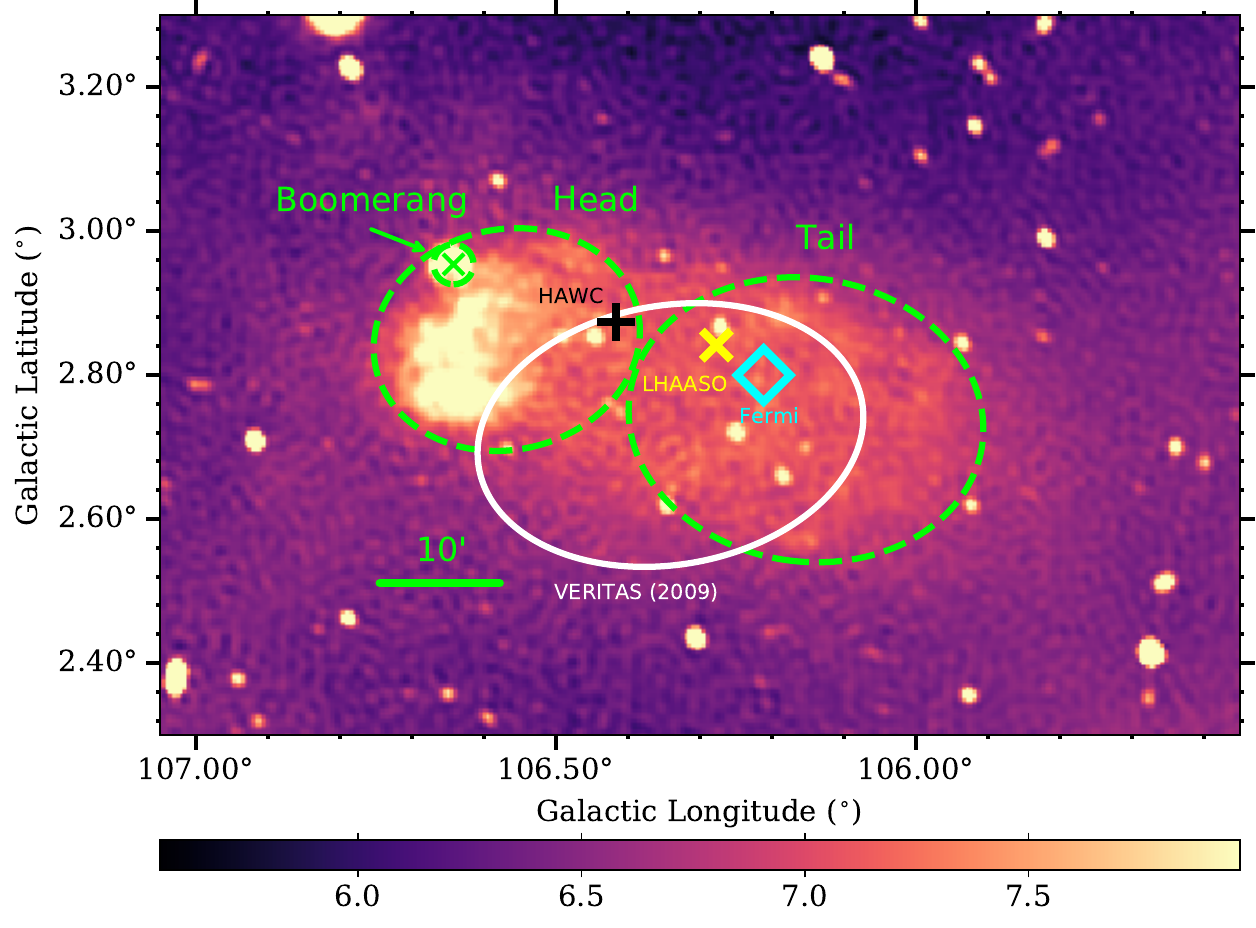}
  \caption{CGPS 1420 MHz radio temperature brightness map [K] of the SNR~G106.3$+$2.7 region with the head, tail and PWN indicated by green dashed lines. The pulsar location is marked by the green cross. The white ellipse represents the extent of the gamma-ray emission previously detected by VERITAS. The black plus, yellow cross, and cyan diamond represent the centroids of the gamma-ray emission detected by HAWC, LHASSO, and \fermi, respectively.}
  \label{fig:SNRreg}
\end{center}
\end{figure}

The Boomerang region is one of the most remarkable composite SNRs for its complex multi-wavelength morphology and the recent detection of gamma rays above 100~TeV indicating it to be a PeVatron candidate.  
Its large-scale radio emission (G106.3$+$2.7) consists of a compact boomerang-shaped nebula around the radio pulsar PSR~J2229$+$6114 and cometary structure extending toward the southwest. The radio source G106.3$+$2.7 was first identified as a SNR by \cite{Joncas&Higgs1990} following the Dominion Radio Astrophysical Observatory (DRAO) survey of the northern Galactic plane. Using further DRAO observations in the 408 MHz and 1420 MHz continuum bands, \cite{PineaultJoncas2000} discerned two distinct regions of SNR~G106.3$+$2.7, labeled the head and the tail (See Figure \ref{fig:SNRreg}). The head region is characterized by its higher surface brightness and flatter spectral index in comparison to the elongated tail region. Using VLA observations at 20 and 6 cm, as well as ROSAT and ASCA observations, \cite{Halpern2001a} identified a compact radio and X-ray source in the northeast area of the SNR~G106.3$+$2.7 head region and suspected it to be a pulsar with a corresponding PWN. The radio and X-ray detections of a 51.6 ms pulsation from the pulsar, now known as PSR J2229+6114, confirmed this hypothesis \citep{Halpern2001b}. Further radio and X-ray timing studies of the pulsar led to determining a spin-down power of $2.2 \times 10^{37}$ erg s$^{-1}$ and a characteristic age of $\sim$10 kyr \citep{Halpern2001b}. A compact PWN with a $r\sim100$\asec\ extension was detected in the radio band and was suggested to be associated with SNR~G106.3$+$2.7 based on the subsequent measurement of the same peak H I velocity from the compact Boomerang nebula and the head region \citep{Kothes2001}. While SNR~G106.3$+$2.7 has been labeled as an SNR, no thermal X-ray emission is reported anywhere in the Boomerang complex, and no large-scale radio morphological features are evident that might suggest the supernova blast wave. The larger-scale integrated radio spectral index is $-0.61$ \citep{Kothes2006}, while that of the PWN alone is $\sim 0$ \citep{Halpern2001b}, suggesting a shock acceleration source for the larger scale electrons, but there is no edge brightening apparent in any location.

It has been hypothesized that the Boomerang's shape could be caused by a bow-shock between \psr\ and its surrounding medium. However, this was deemed unlikely, as simple modelling of the system under this assumption resulted in a supernova explosion energy far below anything ever recorded; the pulsar also does not lie at the apex of the bow structure \citep{Kothes2001, Kothes2006}. In contrast, based on the boomerang-like radio morphology as well as its proximity to the northeast boundary of SNR~G106.3$+$2.7, it was postulated that the PWN had been crushed by a SNR reverse shock \citep{Kothes2001, Kothes2006}. Further observations of the Boomerang region with the Effelsberg 100-m telescope led to a hypothesis that an interaction with the SNR reverse shock could also account for the low radio luminosity of the PWN with respect to the spin-down power \citep{Kothes2006}. Furthermore, a radio spectral break observed between 4 and 5 GHz was attributed to synchrotron cooling. Under this assumption, \citet{Kothes2006} suggested that the PWN B-field is 2.6~mG and that the PWN was crushed by the SNR reverse shock 3900 years ago. A more recent study based on a model of the diffusion of the relativistic electrons injected into the PWN and X-ray radial profile suggested that the PWN B-field is 140~$\mu$G \citep{Liang2022}.

In the X-ray band, the pulsar, its compact nebula ($r \sim 30$\asec) and diffuse emission over $r \simlt 100$\asec\ were  detected by two \chandra\ observations (17 and 94 ksec) in 2001--2002  \citep{Halpern2001b}. \xmm\ and \suzaku\ observations revealed more extended diffuse X-ray emission from the head and tail region \citep{Ge2021, Fujita2021}. 
The long \chandra\ observation in 2002 unveiled a point source at the pulsar position, an incomplete torus of $r\sim10$\asec\ and a jet-like feature. These X-ray features resemble those of the Vela PWN whose pulsar's motion is aligned with its X-ray jet \citep{Halpern2002}. The \chandra\ ACIS image of the Boomerang PWN was fit by a 3D torus model \citep{Ng2004}. The brighter side of the torus (west; see Figure~\ref{fig:ChandraImage}) is due to Doppler boosting of mildly relativistic magnetic hydrodynamic (MHD) outflow from the termination shock. The best-fit torus model predicts that the pulsar should be moving along the spin-axis (i.e. the jet direction toward the northwest). The prediction that the pulsar is moving toward the northwest does not seemingly agree with the tail morphology of SNR~G106.3$+$2.7. However, the recent numerical studies studying the evolution of PWN while interacting with a host SNR show that the PWN's morphology depends on both the pulsar's proper motion and the region's density gradient \citep{Kolb2017}.

Gamma-ray emission has been observed in the SNR~G106.3$+$2.7 region from GeV up to few hundreds of TeV energy range. The Large Area Telescope on board of Fermi gamma-ray space telescope (\fermi)  detected GeV emission coincident with PSR J2229+6114, which was also associated with EGRET source 3EG J2227+6122 \citep{Abdo2009a,Hartman1999}. Gamma-ray pulsations were observed above 0.1~GeV, confirming GeV emission originates from PSR J2229+6114 \citep{Abdo2009b}. Using a collection of \fermi\ data, \cite{Xin2019} identified emission between 3-500 GeV coincident with the tail region with a source radius $0.25^{\degree}$. The Very Energetic Radiation Imaging Telescope Array System (\veritas) detected TeV emission from the tail region and found that the centroid of the TeV source overlaps with $^{12}$CO cloud $J=1\rm{-}0$ emission \citep{Acciari2009}. More recently, the MAGIC collaboration reported TeV detection of the head region as well \citep{Oka2021}.  Gamma-ray emission with energies higher than 100 TeV was detected by HAWC \citep{Albert2020}, Tibet AS $\gamma$ \citep{Amenomori2021} and LHAASO  \citep{Cao2021}; the UHE source is coincident with the \veritas\ and \fermi\ tail region sources as well as PSR J2229+6114 \citep{Albert2020}. The UHE detection identified the Boomerang region as a PeVatron candidate, but its origin is still debated between the leptonic and hadronic cases associated with the Boomerang PWN and the SNR interaction with molecular clouds, respectively \citep{Ge2021, Bao2021, Fujita2021, Breuhaus2022, Liu2022}. Various high energy emission centroids/extents are depicted over the 1420 MHz radio temperature brightness map of SNR~G106.3$+$2.7 in Figure \ref{fig:SNRreg}.
\begin{deluxetable*}{ccc}[]
\tablecaption{Boomerang distance estimates}
\tablecolumns{3}
\tablehead{ \colhead{d}   & \colhead{method} & \colhead{citation} } 
\startdata   
0.8 kpc & H I radial velocities & \cite{Kothes2001} \\
3 kpc & column density ($N_H$) & \cite{Halpern2001a} \\
5-7.5 kpc & dispersion measurement & \cite{Yao17, Abdo2009b} \\
12 kpc & H I radial velocities & \cite{PineaultJoncas2000} \\
\enddata
\label{tab:distances}
\end{deluxetable*}

The distance to the Boomerang complex is unusually poorly determined, even among supernova remnants. A list of the various distance measurements is provided in Table 1. \cite{PineaultJoncas2000} reported radio continuum observations with DRAO at 408 and 1420 MHz, as well as H~I observations, in which they identified an absorption feature at $-104$ km s$^{-1}$, giving a kinematic distance of 12 kpc. This would put \src\ at a $z$-height of 607 pc above the Galactic plane, with a linear extent of over 200 pc. However, \cite{Halpern2001a} suggested a much closer distance based on the measurement of an absorbing column density. They observed the PWN region with a radio (VLA) and X-ray (ASCA) telescope in order to search for a counterpart of the unidentified EGRET gamma-ray source 3EG J2227+6122. In this study, they obtained an absorbing column $N_H$ of $6.3 \times 10^{21}$ cm$^{-2}$. Since the column density through the entire Galaxy is only $8.4 \times 10^{21}$ cm$^{-3}$ in that direction, they concluded that the PWN was at least 2 kpc away, perhaps much further, and assumed a fiducial distance of 3 kpc. The pulsar discovery \citep{Halpern2001b} reported a DM of $200 \pm 10$ \DM , which, with the Taylor \& Cordes (1993; TC93) electron-density model, implies a distance of 12 kpc \citep{Taylor93}. A revised $n_e$ model, NE2001 \citep{Cordes2002}, gives 7.5 kpc for the same DM value \citep{Abdo2009b}. \citet{Yao17} proposed a new $n_e$ model (YMW16) to estimate the distance to the pulsars using the same DM value. YMW16 estimated the distance to the Boomerang PWN to be 5.037~kpc with an error of 40~\%. This error is larger than 20~\%, the threshold YMW16 considered for their model estimation to be satisfactory. \citet{Yao17} also notes that the distance to the Boomerang PWN showed the largest impact due to the Galactic warp. A distance of 3 kpc would imply, from TC93, a DM of only 75 \DM. \cite{Kothes2001} suggested a very near distance, 800 pc, based on morphological correspondences between the radio continuum image and channel maps of H I and CO from surveys, at velocities of about $-6$~km s$^{-1}$. \cite{Kothes2004} presented a new technique of H I absorption of polarized emission for distance determinations; they pointed to the absence of an absorption feature in the range of $-70$ to $-55$ km s$^{-1}$, which they asserted would be present if \src\ were further away than the Perseus arm at about 3 kpc. 

The ambiguity in the distance of \src\ may be rooted in its relatively high Galactic latitude of $2.9^\circ$.  Over 85\% of the 383 Galactic SNRs in the catalog of \cite{Ferrand2012} are closer to the Galactic plane than this. At 3 kpc, for instance, \src\ has a $z$-height of 150 pc, higher than the H I scale height of the Galactic disk of about 100 pc, and perhaps explaining anomalous H I absorption (or its absence).

The properties of \src\ are extreme at either distance: 0.8 kpc or 12 kpc. At 0.8 kpc, as pointed out by \cite{Kothes2006}, the PWN would have an extremely low ratio of radio luminosity to pulsar ${\dot E}$. Additionally, the H I column density from X-ray observations of $6.3 \times 10^{21}$ cm$^{-2}$ implies a mean volume density of neutral atomic H of 2.6 cm$^{-3}$ between Earth and \src\, which is unrealistically high. At 12 kpc, \cite{Halpern2001b} state that the pulsar would need to be more efficient than the Crab or Vela pulsars at converting spindown luminosity into $> 100$ MeV gamma-rays. 

In this paper, we present a multi-wavelength analysis of the Boomerang PWN region. We shall refer to the compact radio and X-ray source within 100\asec\ of the pulsar as the PWN, and as distinct from the head (scale $\sim 15$\amin) and tail (scale $\sim 30$\amin) regions. Our work is focused on this PWN region, in contrast to the recent publication by \citet{Fang2022} for example, which primarily concerns the larger-scale nebula. Our motivation is to better constrain the characteristic properties and gain further insight on the formation of the Boomerang PWN. In doing so, we gain a better understanding of the Boomerang PWN's relationship to the high-energy emission coincident with SNR~G106.3$+$2.7, mostly confined to the SNR's tail region. We begin by describing the archival \chandra\ and new \nustar\ X-ray observations and our timing, imaging and spectral analysis (\S\ref{sec:x-ray obs}). In \S\ref{sec:Fermi} and \S\ref{sec:VERITAS} we describe the gamma-ray observations of the Boomerang PWN region and analysis of the corresponding data from \fermi\ and \veritas, respectively. We then combine the multi-wavelength spectral data of the Boomerang PWN and explore various models that could describe the PWN's emission through SED fitting (\S\ref{sec:SED}). 
For the SED models, we consider the two most extreme source distances, 0.8 and 7.5 kpc, from the Table 1. (From the two distance measurements based on H I radial velocity measurements, we chose 0.8 kpc over 12 kpc as it is more the most recent estimation.) In the end, we determine the source distance from the SED and X-ray morphology Tanalysis of the Boomerang PWN region,
within 100\asec\ of the pulsar; We do not consider the emission on
larger scales. Finally, we discuss the results from our X-ray and multi-wavelength analysis and constrain the PWN B-field  (\S\ref{sec:Discussion}). We contemplate the current evolution phase of the Boomerang PWN and examine its relation to the high-energy emission coincident with SNR~G106.3$+$2.7. We summarize our results in \S\ref{sec:Conclusions}. 

\section{Observations and data analysis}

We present X-ray, GeV (\fermi) and TeV (\veritas) gamma-ray observations of the Boomerang PWN in the following sections. We performed X-ray analysis of the 2002 \chandra\ and 2020 \nustar\ observation data (\S\ref{sec:x-ray obs}). \fermi\ and \veritas\ data analysis was confined only to the Boomerang PWN region, and the source extraction region varied according to the point spread function (PSF) of each telescope (\S\ref{sec:Fermi} and \ref{sec:VERITAS}). All errors are given to the $2\sigma$ confidence level unless explicitly stated otherwise. 

\subsection{X-ray observations}
\label{sec:x-ray obs}

\chandra\ observed the Boomerang region with its CCD ACIS-I array on 2002 March 15 (Obs ID 2787, PI Halpern) for $\sim94$ ks of exposure. The observation files were processed and analyzed using the tools in \texttt{CIAO v4.13}. \nustar\ observed \psr\ and its PWN on 2020 September 21 (ObsID 40660001002) for a total of ~45 ks of exposure. Data analysis was conducted using the \texttt{NuSTARDAS v2.0.0} sub-package within \texttt{HEASOFT v6.28}. \texttt{CIAO v4.13} was also used for its image modelling and fitting application (\texttt{SHERPA}). 

The Boomerang PWN region is composed of four components: (1) the pulsar (\psr), (2) a torus--jet feature which represents the termination shock region  \citep[$r\sim10$\asec; ][]{Ng2004}, (3) X-ray PWN ($r \sim 30$\asec), and (4) diffuse X-ray emission ($r \sim 100$\asec).  These X-ray features are resolved by \chandra\  as shown in the \chandra\ image in the 0.5--8.0 keV band (Figure \ref{fig:ChandraImage}). Since \nustar\  (with 58\asec\ half-power diameter) cannot spatially resolve the pulsar from the extended X-ray emission, we first performed timing analysis on the \nustar\ data in order to remove the pulsar emission (\S\ref{sec:Timing}). We then present \nustar\ imaging analysis in different energy bands in comparison with the high-resolution \chandra\ images below 8 keV (\S\ref{sec:Imaging}).  In \S\ref{sec:spectroscopy}, we  analyze X-ray spectral data of the PWN by excising the pulsar emission spatially (for \chandra) and by selecting a phase interval for the off-pulse component (for \nustar). 

\subsubsection{\nustar\ timing analysis}
\label{sec:Timing}
\begin{figure}[b!]
\begin{center} 
  \includegraphics[width=0.85\textwidth]{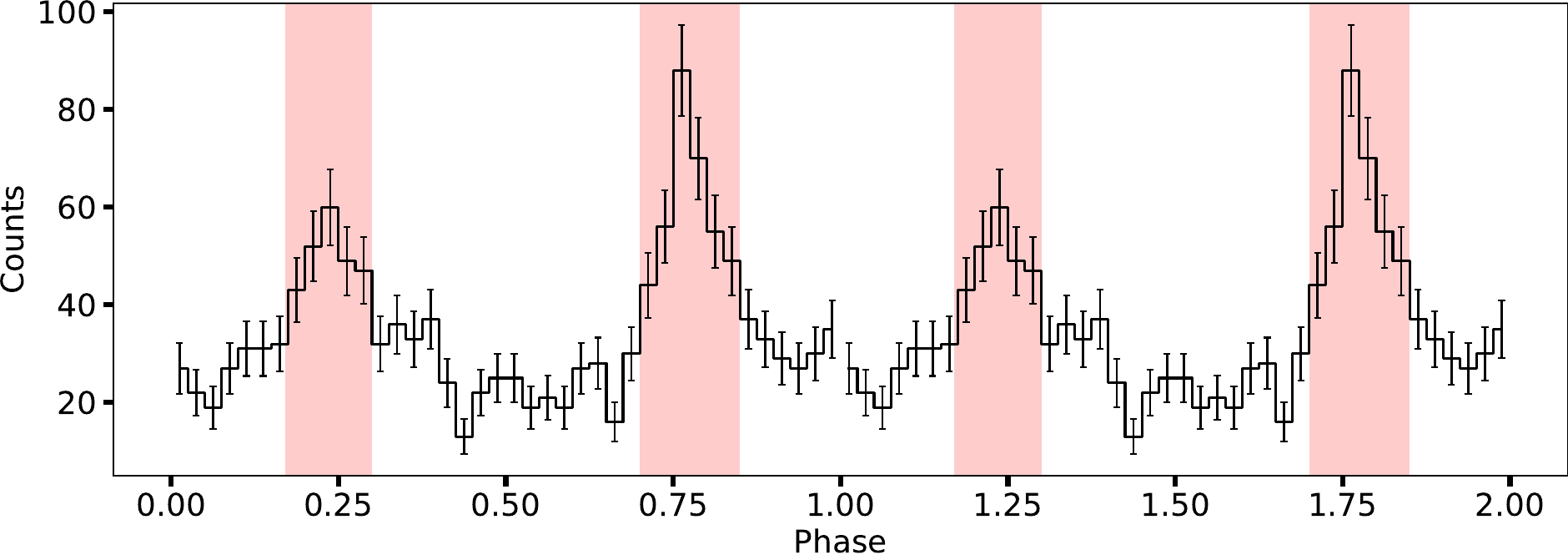}
  \caption{\nustar\ 3--20 keV folded lightcurve of \psr. The pulsed phase ranges excised from our PWN imaging and spectral analysis are demarcated by the red regions.}
  \label{fig:lightcurve}
\end{center}
\end{figure}
\begin{figure*}[t!]
\begin{center} 
    \includegraphics[trim=3cm 0 0 0, width=0.55\textwidth]{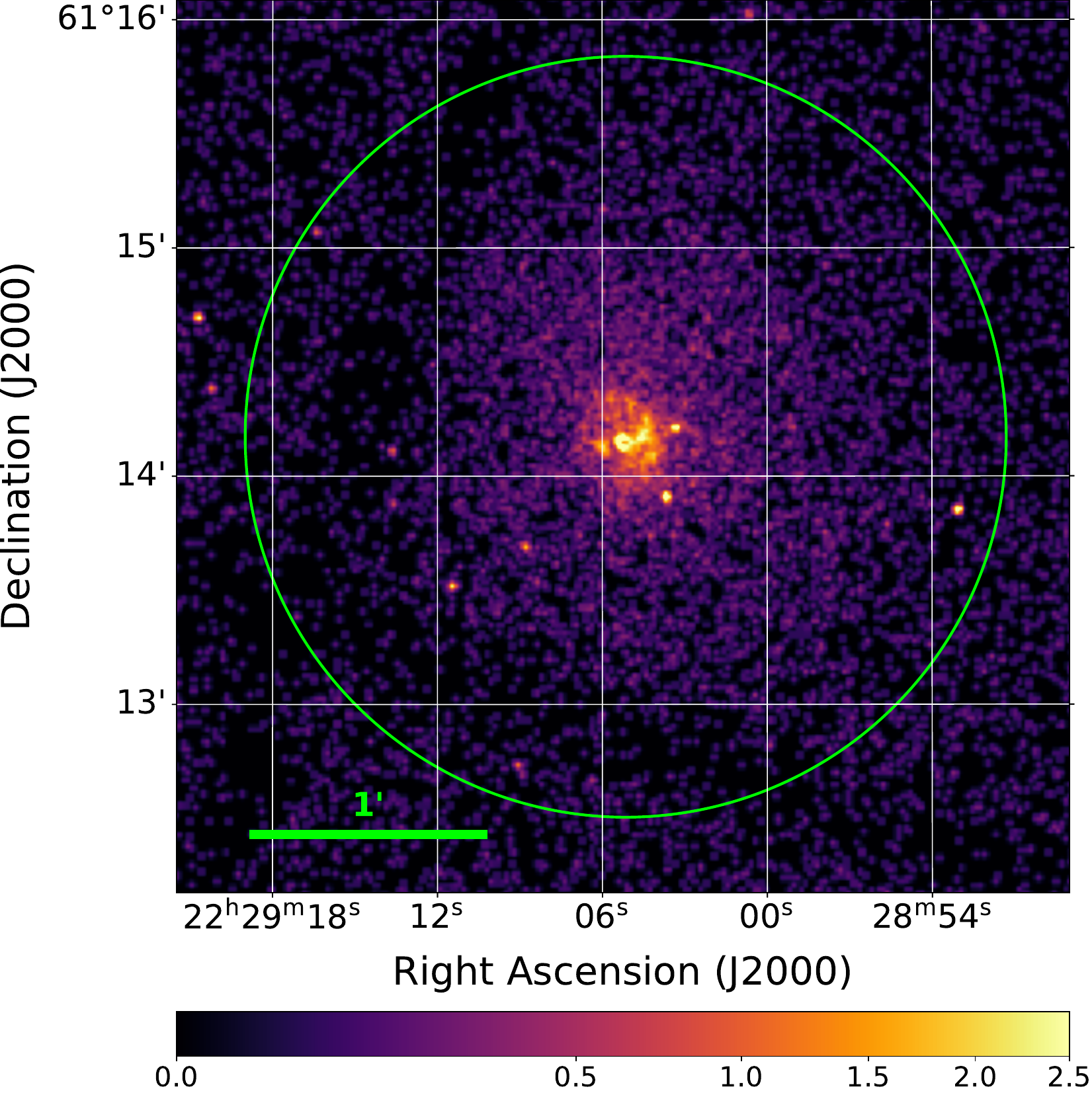} \\
    \vspace{0.5 cm}
    \includegraphics[trim=3cm 0 0 0, width=0.55\textwidth]{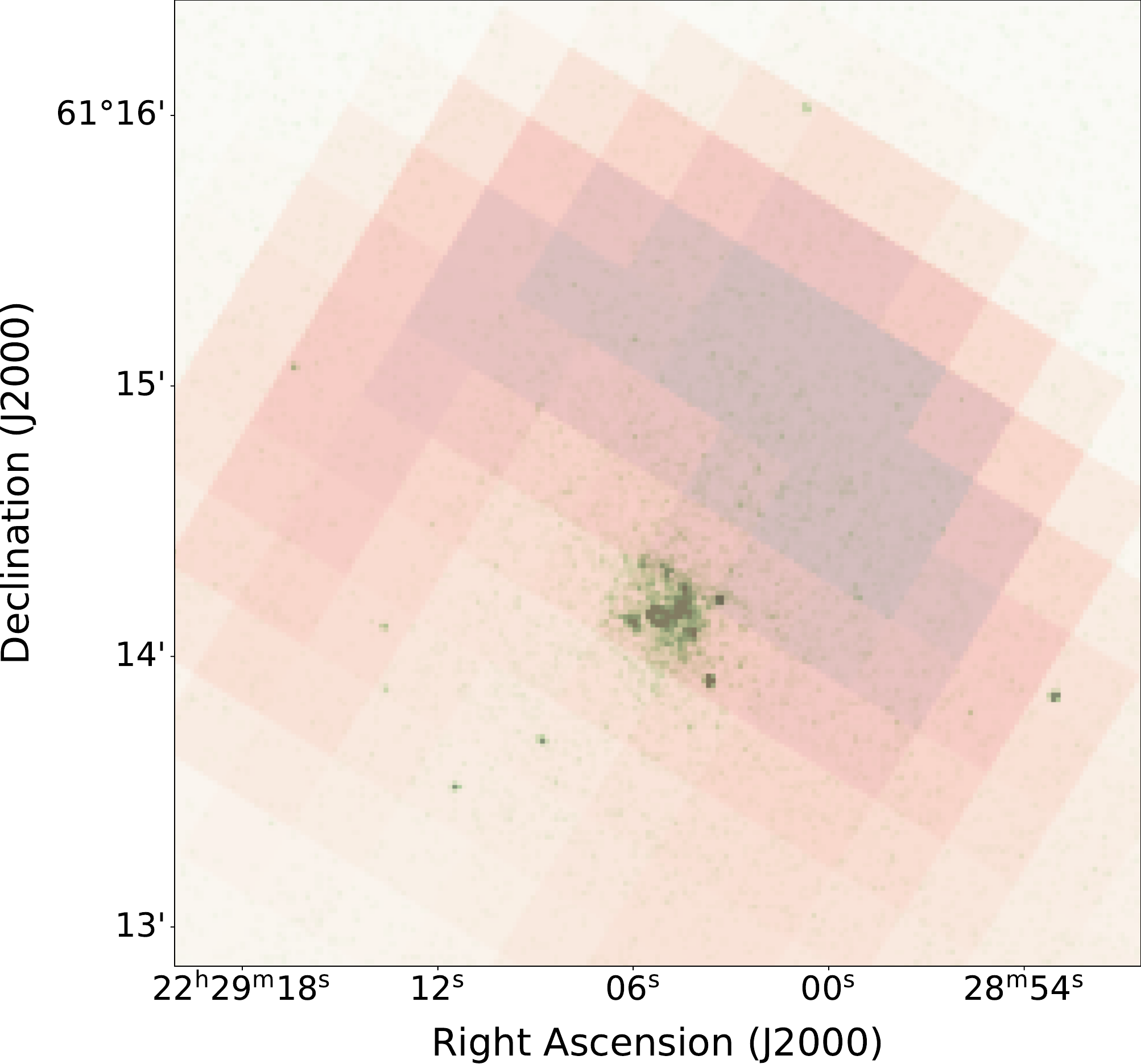}
    \caption{
    (top) \chandra\ 0.5-8 keV image of the Boomerang PWN, smoothed to $\sigma=\sim0.5\asec$ (1 pixel). The color scale is in units of counts per pixel. A $r=1\asec$ circular region around the pulsar was excised in order to accentuate the PWN features. The solid green circular region is a $r=100$\asec\ circular region around the pulsar position, used for spectral extraction.
    (bottom) 0.5-8 keV \chandra\ image (same as from top panel) is shown here in green, overlain with the CGPS 1420 MHz radio temperature brightness map of the Boomerang region in red.
    }
    \label{fig:ChandraImage}
\end{center} 
\end{figure*}
We found that X-ray emission from the Boomerang  pulsar + nebula system was detected up to 20 keV. We limited our timing analysis to the 3--20 keV band. The \nustar\ telescope consists of two focal plane modules (FPMA and FPMB) which are described in detail in \cite{Harrison2013}. We determined the source centroid for both FPMA and FPMB images using DS9's centroid function. We then applied barycentric correction to the event files using \texttt{barycorr}, and extracted source events from a $r=30$\asec\ circular region around the source centroid. Using the \texttt{Stingray} X-ray timing analysis package \citep{matteo_bachetti_2021_4881255}, we applied the $Z^{2}$ algorithm with two harmonics in order to search the combined event times from both focal plane modules for pulsations. A strong periodic signal with $P = 51.671495^{+1\times10^{-6}}_{-3\times10^{-6}}$ ms ($3\sigma$ error bars) was detected with $Z_{2}^{2} = 174$. This period differs slightly from that reported by \citet{Halpern2001b} after accounting for the $\dot{P}$ quoted in the same paper, $P = 51.67199357(0)$, indicating an undetected timing glitch. Folding upon the measured period, we generated a pulse profile in the 3--20 keV band  (Figure \ref{fig:lightcurve}). The on-pulse component is clearly identified as an asymmetric double-peak in the folded lightcurve, consistent with the pulse profiles measured by \citet{Halpern2001b} and \citet{Kuiper2015}. We considered our baseline off-pulse component to be around 25 to 30 counts per bin. As a conservative estimate for the pulsed emission, we only excised the clear peaks significantly above the baseline. After calculating a phase value for each photon event, we removed the on-pulse events between $\phi = 0.17$ and 0.30, as well as $\phi = 0.70$ and 0.85 using \texttt{extractor} for subsequent spectral and imaging analysis of the nebular emission. The effect of any leftover pulsar component post phase extraction is negligible, as we determine in Section \ref{sec:spectroscopy}.
\subsubsection{\nustar\ and \chandra\ imaging analysis}
\label{sec:Imaging}
\begin{figure*}[b!]
\begin{center} 
    \includegraphics[trim=3cm 0 0 0, width=0.60\textwidth]{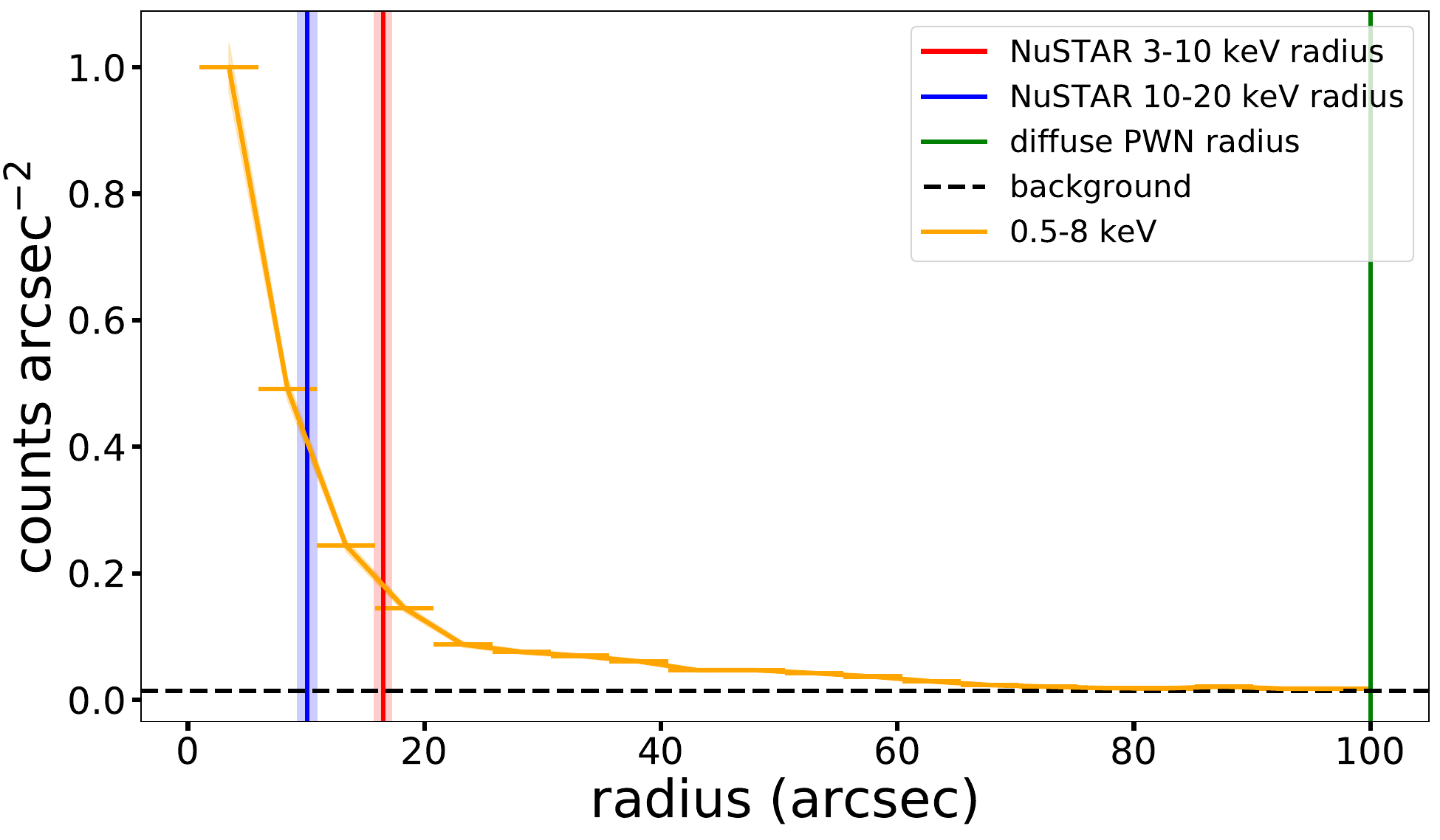}
    \caption{
    \chandra\ 0.5--8 keV radial profile of the Boomerang PWN. The vertical lines indicate the PWN radius as measured from the diffuse X-ray emission (green), \nustar\ 3--10 (red) and 10-20 keV (blue) energy bands. The 1 sigma errors are indicated by the shaded regions. The error bars designate the $\sim$5\asec\ radial bin widths.
    }
    \label{fig:ChandraRadial}
\end{center}
\end{figure*}

The \chandra\ observations detected the Boomerang PWN extending to the bounds of its radio emission \citep[$r \approx 100\asec$; ][]{Halpern2001b}. We thus considered $r=100$\asec\ as the outermost  boundary of the X-ray nebula for \nustar\ imaging analysis using the phase-resolved event files after excising the pulsed emission. The broad bandwidth of \nustar\ allows us to compare the X-ray image of Boomerang in different energy ranges. We performed energy-resolved imaging analysis in a ``soft'' band (3--10 keV) and ``hard'' band (10--20 keV). For each of the FPMA and FPMB data, we corrected the positional offsets between the \nustar\ source centroid (measured in the on-pulse images) and the pulsar position measured by \chandra. 

\begin{figure*}[]
    (a) {{\includegraphics[width=0.40\textwidth]{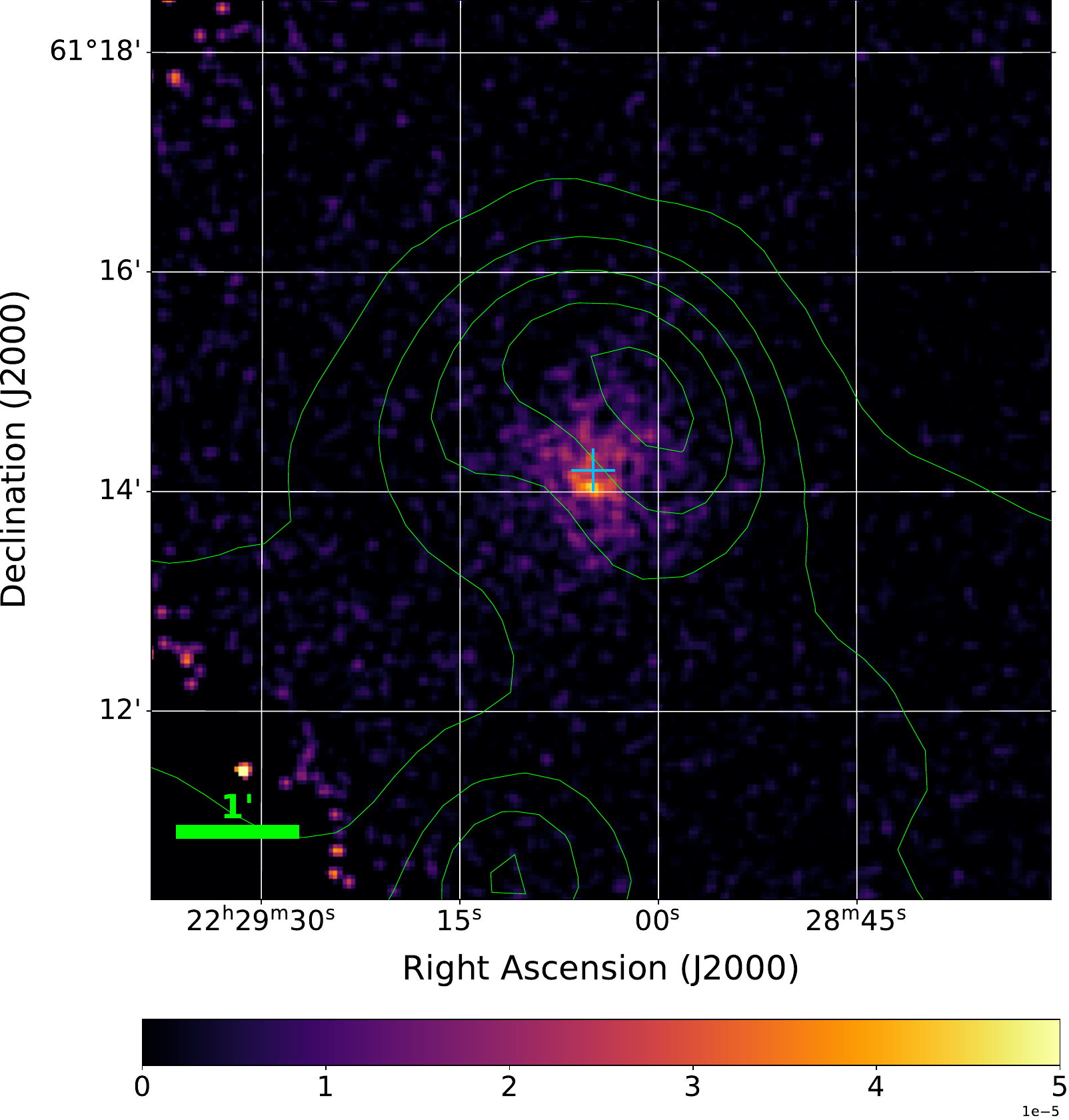} }} 
    \qquad 
    (b) \includegraphics[width=0.40\textwidth]{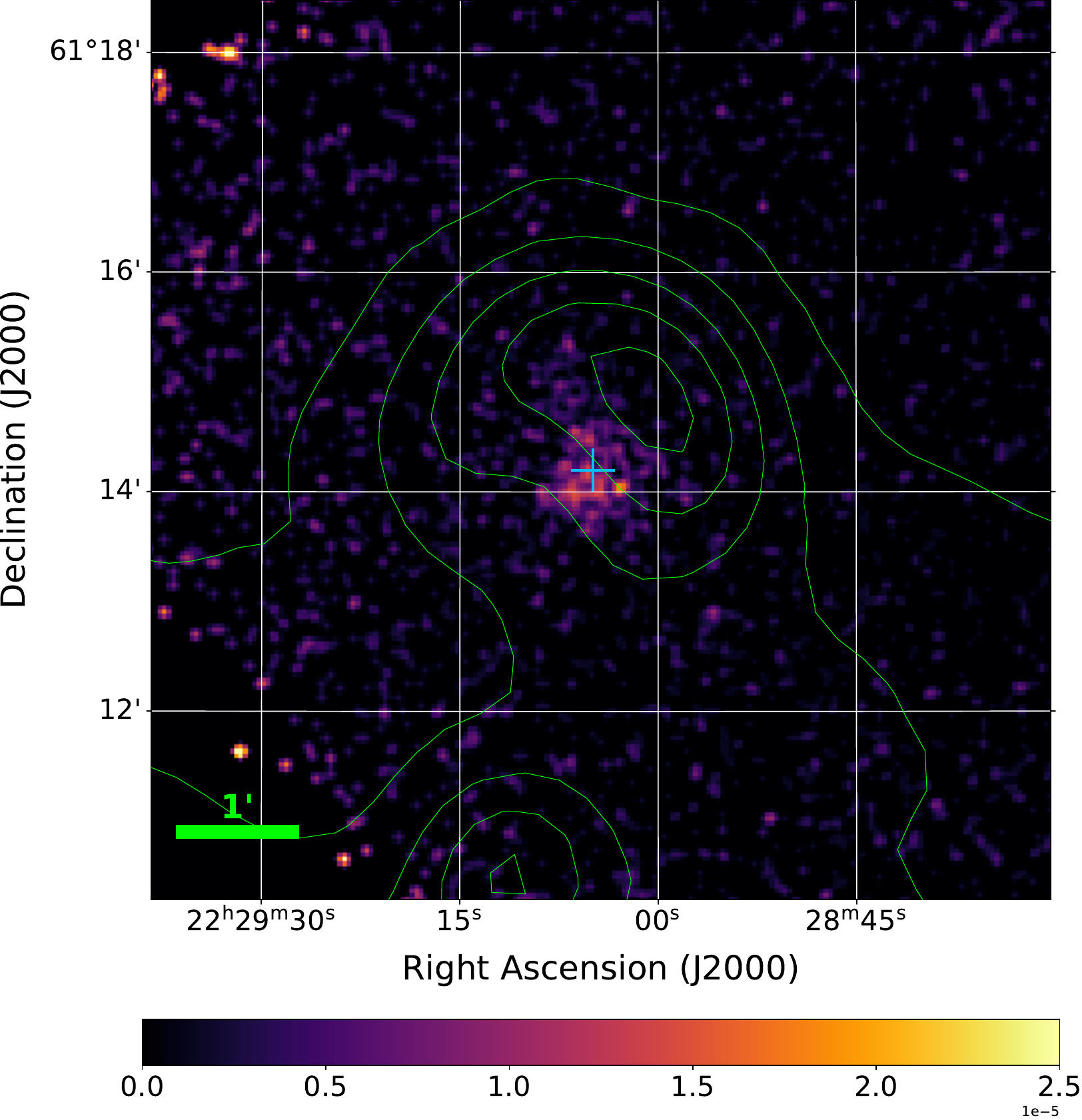} 
    \caption{
    \nustar\ 3--10 keV (a) and 10--20 keV (b) background subtracted flux images of the Boomerang PWN [10$^{-5}$ counts s$^{-1}$ cm$^{-2}$]. 1420 MHz radio contours are shown in green and the position of PSR~J2229$+$6114 is marked by the blue cross.
    }
    \label{fig:NuSTARImage}
\end{figure*}

The event files for both FPMA and FPMB were split into the soft and hard bands with \texttt{extractor}. Exposure maps were created with \texttt{nuexpomap} for each event file with vignetting effects at 6.5 and 15 keV for the soft and hard band, respectively. The FPMA and FPMB event files were combined for each energy band, and the same was done for the exposure maps. For the purpose of smoothing out spurious features near the detector edges, the summed \nustar\ images were convolved with a Gaussian kernel of $\sigma = 2.46$\asec\ (corresponding to the \nustar\ pixel size) before being divided by the corresponding exposure maps \citep{Nynka2014}. The above process produced an exposure corrected mosaic flux image for each energy band. In each energy band, we calculated the background level using a region to the northeast of Boomerang, avoiding the diffuse X-ray emission detected by \cite{Ge2021}. The resultant 3--10 keV and 10--20 keV background subtracted flux images of the Boomerang PWN are shown in Figure \ref{fig:NuSTARImage}.

\begin{figure}[t!]
\begin{center} 
  \includegraphics[width=0.60\textwidth]{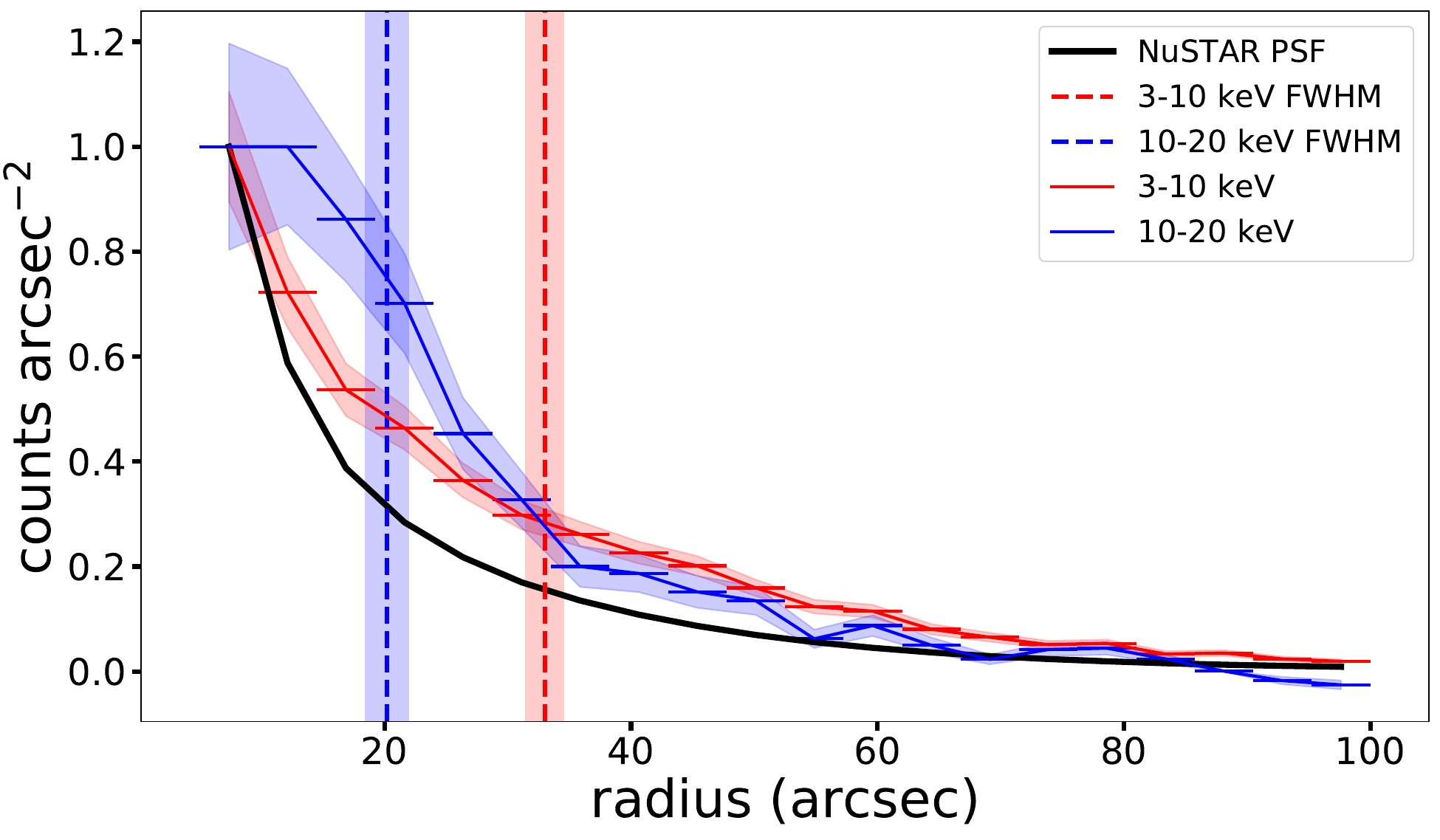}
  \caption{\nustar\ X-ray radial profiles of the Boomerang PWN in the soft (red) and hard (blue) bands, compared to the \nustar\ PSF. The vertical lines indicate the SHERPA best fit 2D-Gaussian FWHM in each energy band. The 1 sigma errors are indicated by the shaded regions. The horizontal bars designate the 4.75\asec\ radial bin widths.}
  \label{fig:radialprofile}
\end{center}
\end{figure}

In order to characterize the Boomerang PWN emission, we compared radial profiles around the pulsar position in the two energy bands, as well as the \nustar\ PSF for determining a source extension. A set of 20 annuli between $r_{\rm in} = 5\asec$ and $r_{\rm out} = 100\asec$ were centered on the source centroid of the mosaic image. The radial profiles for each energy band were extracted from these annuli and normalized so that the brightness was set to 1 at $r=0$. The same set of annuli was used to create a normalized radial profile of the \nustar\ 8--12 keV PSF to serve as a point source template. Since we found that the \nustar\ PSF varies insignificantly in 3--20 keV, we chose 8--12 keV to produce the radial profile representative for our case. The soft, hard, and PSF radial profiles are shown in Figure \ref{fig:radialprofile}.


The background subtracted source profiles, as shown in Figure \ref{fig:radialprofile}, are extended above the \nustar\ PSF up to $r\sim100$\asec\. The radial profiles in both the soft and hard bands appear more extended than the \nustar\ PSF profile.  Furthermore, the soft band exhibits a slightly wider radial profile than that of the hard band. To determine the size of the nebula in each energy band more quantitatively, we fit the \nustar\ images using \texttt{SHERPA}. We modelled the X-ray source as a 2D-Gaussian and included a constant background level. The source model was convolved with the \nustar\ PSF and then fit to the \nustar\ image data. After taking into account the telescope dithering, we produced the effective \nustar\ PSF data in 4.5--6 keV and 12--20 keV for the soft and hard bands, respectively \citep{Nynka2014}. The fit yielded a full width at half maximum (FWHM) of  $33\pm2$\asec\ for the soft band and $20\pm2$\asec\ for the hard band; the errors represent the 1 sigma confidence intervals (see Figure \ref{fig:radialprofile}).

For the purpose of illustrating the various regions of interest in the Boomerang PWN and comparing with the \nustar\ images, we created a high-resolution radial profile of the \chandra\ 0.5--8 keV image shown in Figure \ref{fig:ChandraRadial}. To produce this radial profile, we generated a set of 20 annuli around the pulsar position from $r_{\rm in} = 1\asec$ (to mask out the pulsar emission) to $r_{\rm out} = 100\asec$ (i.e the boundary of the X-ray nebular emission). The profile produced from these annuli was normalized, and the background surface brightness was extracted from the same region used for background spectra extraction. The radius of the diffuse X-ray extent measured by \cite{Halpern2001b} and the \nustar\ 3--10 and 10-20 keV PWN radii are plotted over the resulting radial profile in Figure \ref{fig:ChandraRadial}. The FWHM of the soft and hard X-ray emission detected by \nustar\ ($r=33$\asec\ and 20\asec) corresponds to the inner PWN where most of the X-ray nebular emission is concentrated.

\begin{figure*}[b!]
    \centering
    (a) {{\includegraphics[width=0.45\textwidth]{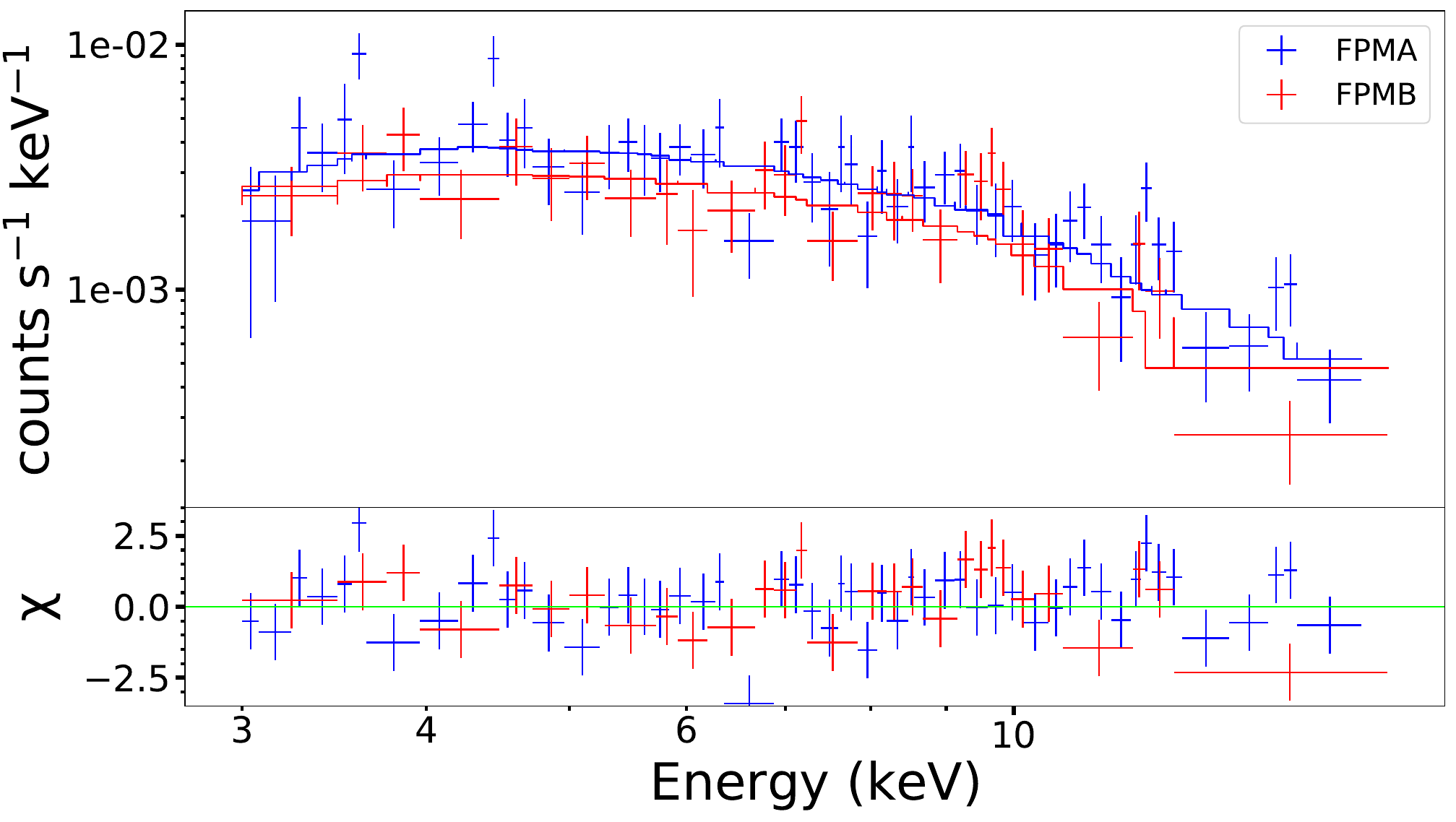} }} 
    (b) \includegraphics[width=0.45\textwidth]{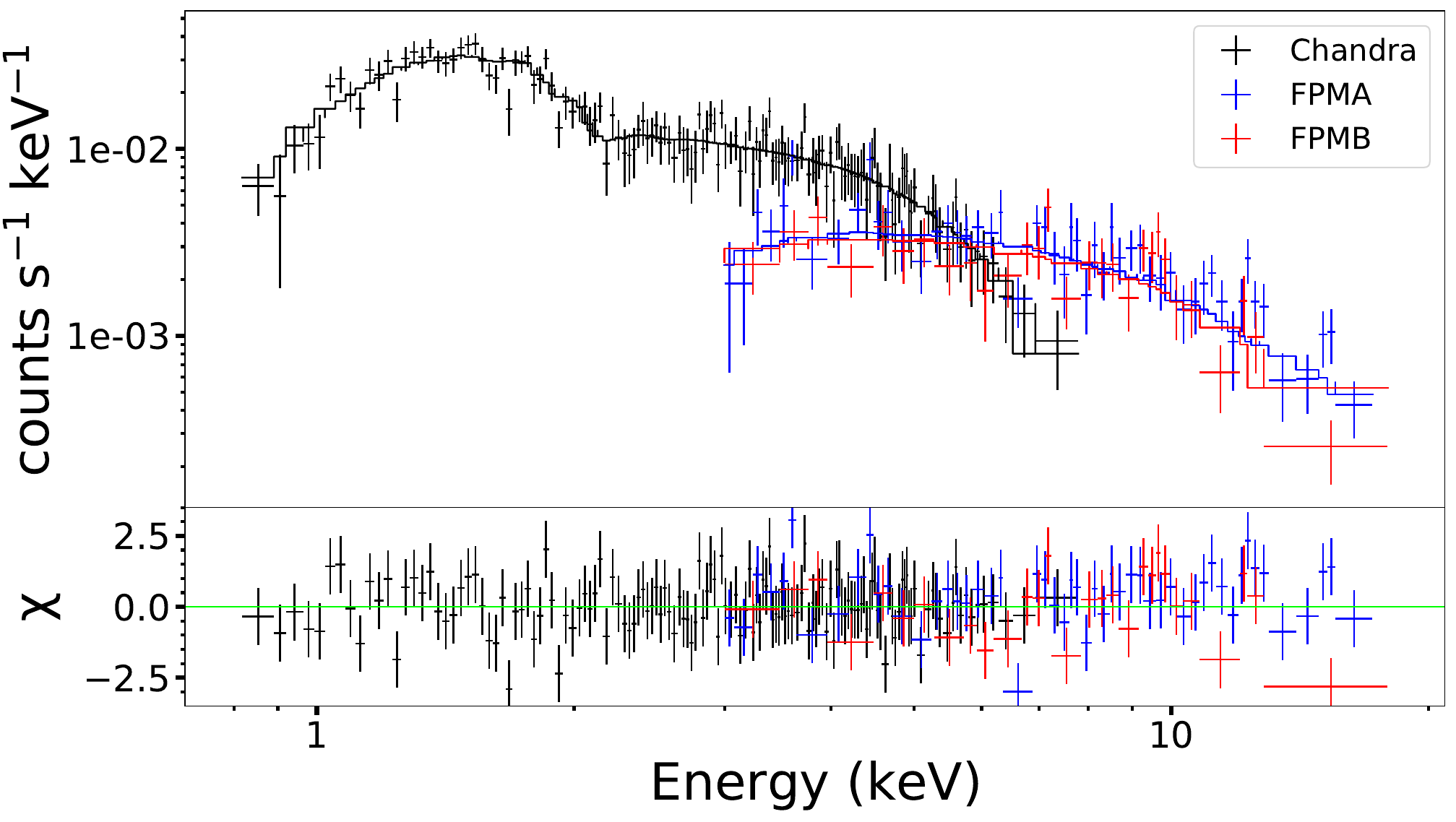} 
    \caption{\nustar-only spectra (a) and \nustar--\chandra\ joint spectra (b) fit to an absorbed power-law model.
    }
    \label{fig:spectra}
\end{figure*}
\begin{deluxetable*}{lccc}[b!]
\tablecaption{Boomerang PWN spectral fitting parameters. All errors are given to the 90\%\ confidence level.}
\tablecolumns{4}
\tablehead{ \colhead{Parameter}   & \colhead{\nustar}  & 
\colhead{\nustar\ + \chandra}  & 
\colhead{\chandra}}
\startdata   
Model & \texttt{tbabs*const*pow} & \texttt{tbabs*const*pow} & \texttt{tbabs*pow}\\
$N_H$ [$10^{22}$ cm$^{-2}$] & 0.89 (frozen) & 0.89 (frozen) & 0.89$^{+0.15}_{-0.14}$\\
$\Gamma_X$ & 1.52 $\pm$ 0.15 & 1.52 $\pm$ 0.06 & 1.52$^{+0.13}_{-0.12}$\\
Flux normalization\tablenotemark{a} & 1.89$^{+0.63}_{-0.49}$ & 1.95 $\pm$ 0.11 & 1.94$^{+0.33}_{-0.28}$\\
Cross-normalization & 0.85$^{+0.11}_{-0.10}$ & 0.92$^{+0.09}_{-0.08}$ & --- \\
$\chi^2_{\nu}$ (d.o.f) & 1.21 (79) & 0.95 (226) & 0.79 (141)\\
$F_X$ (0.5--10 keV)\tablenotemark{b} & 0.98 $\pm$ 0.06 & $1.10\pm0.03$ & 1.15 $\pm$ 0.04\\
$F_X$ (10--20 keV)\tablenotemark{b} & 0.70 $\pm$ 0.04 & --- & --- \\
\enddata 
\tablenotetext{a}{Flux normalization at 1 keV [$10^{-4}$ photons\,cm$^{-2}$\,s$^{-1}$\,keV$^{-1}$]. For the \nustar\ and joint spectral fit, the flux normalization corresponds to the FPMA and \chandra\ spectra, respectively.} 

\tablenotetext{b}{Absorbed X-ray flux in $10^{-12}$ \fluxcgs.} 

\label{tab:xspecfit}
\end{deluxetable*}

\subsubsection{\chandra\ and \nustar\ spectral analysis}
\label{sec:spectroscopy}
In this section, we present our \chandra\ and \nustar\ spectral analysis of the entire PWN region from $r=100$\asec. This region is in accordance with the Boomerang radio PWN, thus it is appropriate for subsequent SED studies (e.g., \S\ref{sec:NAIMA}). Using the \texttt{CIAO specextract} script, the \chandra\ spectrum was extracted from the $r=100$\asec\ circular region around the pulsar position, excluding the pulsar emission in $r<1$\asec. Increasing the exclusion region to $r<3$\asec~ had no significant effect on our spectral fits, suggesting that $r<1$\asec~ is sufficient for excising the pulsed emission component. A source-free region to the immediate northwest of the source extraction region was used for extracting background spectra. In contrast to the \chandra\ spectral analysis, we performed \nustar\ spectral analysis on the off-pulse events of the Boomerang region. We used a $r=100$\asec\ circular source region around the emission centroid (which was previously adjusted to the pulsar position in each module data). The \nustar\ response matrix (RMF) and effective area (ARF) were created using \texttt{nuproducts} for an extended source option. We produced \nustar\ background spectra in two different approaches by modelling the background using \texttt{nuskybkg} \citep{Wik2014} as well as by extraction from an off-source region with \texttt{nuproducts}. In the former method, we modeled background spectra with \texttt{nuskybkg} by fitting actual background spectra extracted from multiple source-free regions across the \nustar\ detector chips for both modules. In the latter method, a source-free region on the same detector chip as the source was used for generating background spectra. In both cases, background regions were selected to avoid the additional diffuse non-thermal X-ray emission in the head region \citep{Ge2021}. We found that fitting the source spectra with the background spectra produced by either method yielded statistically identical results. Hereafter we present our \nustar\ spectral analysis results with the standard background extraction using \texttt{nuproducts}.

The \nustar\ and \chandra\ spectra were adaptively binned to 2.5 and 2.0 $\sigma$ over background counts, respectively. In order to determine the hydrogen column density accurately, we fit the 0.5--8.0 keV \chandra\ spectrum with an absorbed power-law model in \texttt{XSPEC v12.12.0}. The fit was first carried out using the default abundance data \citep{Anders1989}, resulting in the best-fit column density $N_{\rm H} = 6.2^{+1.0}_{-0.9} \times 10^{21}~\rm{cm}^{-2}$. We then repeated the \chandra\ spectral fit using the Wilms abundance table \citep{Wilms2000}, obtaining  a higher value of $N_{\rm H} = 8.9^{+1.5}_{-1.4} \times 10^{21}~\rm{cm}^{-2}$. 
Radio observations of \psr\ measured its DM to be $(204.97\pm0.02)$ \DM\ \citep{Abdo2009b}. 
Using a linear relation between $N_{\rm H}$ and DM as well as its slope errors \citep{He2013}, we derived $N_{\rm H} = (4.3\rm{-}8.8)\times10^{21}$~cm$^{-2}$. The hydrogen column densities obtained by the \chandra\ observation and estimated from the DM measurement are consistent with each other.  

We proceeded with spectral fitting using this updated $N_{\rm H}$ value found using the Wilms abundance table. The \nustar\ FPMA and FPMB 3--20 keV spectra were jointly fit using an absorbed power-law model with independently varying FPMA and FPMB flux normalization variables, with the hydrogen column density fixed to $N_{\rm H} = 8.9 \times 10^{21}~\rm{cm}^{-2}$. This allowed us to find the ratio between the FPMA and FPMB flux normalization values (i.e. the cross-normalization factor). We also created a joint fit using the \chandra\ 0.5--8.0 keV and \nustar\ 3--20 keV spectra, again with an absorbed power-law model. In this case, FPMA and FPMB were forced to share a flux normalization value, independent of \chandra's. The resulting joint fit measured the photon index to $\Gamma=1.52 \pm 0.06$ with a reduced chi-squared value of $\chi_\nu^{2}=0.95$ (226 dof), consistent with the individual \nustar\ and \chandra\ spectral fit photon indices within error (see Table~\ref{tab:xspecfit}). A broken power-law or an additional spectral component was not statistically required given the goodness-of-fit with a single power-law model. The \nustar\ spectra and \nustar--\chandra\ joint spectra with the best-fit models are presented in Figure \ref{fig:spectra}. The parameters from these fits, as well as from the \chandra\ spectral fit, are listed in Table \ref{tab:xspecfit}. Given that the cross-normalization factor for the joint fit ($C = 0.92^{+0.09}_{-0.08}$) is consistent with 1, no significant X-ray flux change was detected from the PWN between the 2002 \chandra\ and 2020 \nustar\ observations.

In order to confirm that our \nustar\ PWN spectra was not significantly tainted by pulsar emission post phase extraction, we attempted to characterize the off-pulse pulsar emission. While this was difficult because of the low number of pulsed photon counts, we decided upon the following method. In order to characterize the on-pulse pulsar component, we extracted the on-pulse \nustar\ 3--20 keV spectra from the $r=30\asec$ region around the pulsar position and used the corresponding off-pulse spectra as background. We then jointly fit this spectra with the \chandra\ 2--8 keV point source spectra, using a tbabs*(pow+const*pow) model. By setting the constant term to zero for the \nustar\ spectra, to one for the \chandra\ spectra, and allowing the first powerlaw to only fit the \nustar\ spectra, the latter powerlaw characterized the off-pulse pulsar component. We found that the off-pulse pulsar flux component represents only about 10\% of the pulsar emission, and only about 5\% of the off-pulse emission. We therefore forgo redoing the above spectral analysis with a off-pulse pulsar model component; it is negligible.
\subsection{Fermi-LAT data selection and analysis}
\label{sec:Fermi}
\begin{figure}[t!]
\begin{center} 
  \includegraphics[width=0.45\textwidth]{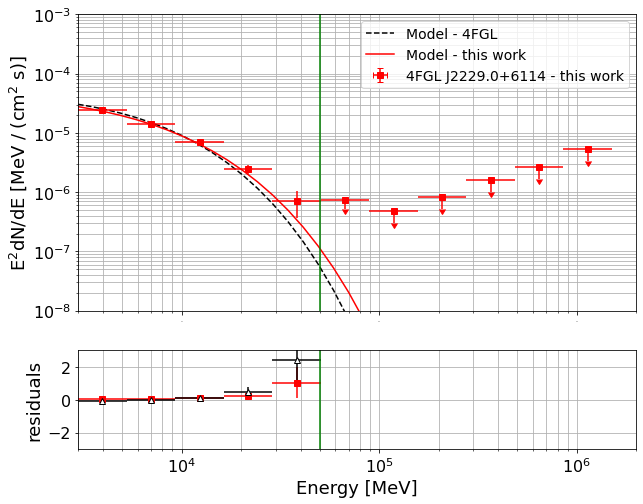}
  \caption{Spectral energy distribution of 4FGL J2229.0+6114. The red points and upper limits were derived from the analysis of LAT data presented in this work. The red solid line indicates the best-fit model to the data points, while the black dashed line is the same spectral model, with the best-fit parameters reported in the 4FGL catalog. The inset reports the residuals as \emph{(data-model)/model}, for the best-fit model derived from this work and the 4FGL catalog one. The green line denotes the threshold at which we separate the pulsar (left) from the nebula (right) emission.}
  \label{fig:fermi_spectrum}
\end{center}
\end{figure}

The Fermi Large Area Telescope (\fermi) is a pair-conversion high-energy (HE) gamma-ray telescope, that can detect gamma rays in the energy range from 20~MeV to above 1~TeV~\citep{2009ApJ...697.1071A}. The presented analysis was performed by means of Fermipy, a python-based package that allows to analyse \fermi\ data with the Fermi Science Tools~\citep{2017ICRC...35..824W}. We used Fermipy version v1.0.1, which is associated to the 2.0.8 version of the Fermi Science Tools.

We selected events with time stamps between MET 239557418 (2008-08-04 15:43:37.000 UTC) and MET 644798253 (2021-06-07 22:37:28.000 UTC), in a 10\degree-wide region around the 4FGL catalog counterpart to PSR J2229+6114, i.e. 4FGL J2229.0+6114. 
The analysis was conducted in the energy range 3~GeV - 2~TeV.

We used P8R3\_SOURCE\_V3 instrument response functions (IRFs) and event type 3 (front and back conversion type). We binned the data using a spatial size of 0.1\degree\ and 8 energy bins per decade. We modeled all the sources within a box of width 20\degree\ that are included in the second release of the 4FGL catalog (4FGL-DR2; \citep{Acero_2016}), along with the isotropic and Galactic diffuse emission (iso\_P8R3\_SOURCE\_V3\_v1 and gll\_iem\_v07).

We first optimized the model by fitting normalization and spectral shape parameters of each source in the region of interest and calculate their Test Statistics (TS = $\mathrm{-2ln(L_{max,0}/L_{max,1})}$ where $\mathrm{L_{max,0}}$ and $\mathrm{L_{max,1}}$ are the maximum likelihoods for a model excluding and including the source, respectively \footnote{https://fermi.gsfc.nasa.gov/ssc/data/analysis/documentation/Cicerone/Cicerone\_Likelihood/Likelihood\_overview.html}), by using the  {\tt gta.optimize} function. We then simplified the model by removing sources that have TS below 4 (i.e. with a detection significance below $\mathrm{\sim 2}$ standard deviations) and number of predicted counts below 1, in order to ease convergence of the fit. Before performing the final fit, we freed sources that have TS above 25 (i.e. with a detection significance of $\mathrm{\sim 5}$ standard deviations or above), within a radius of 5\degree\ from the target position of 4FGL J2229.0+6114, as well as the isotropic and Galactic diffuse emission components. We also modelled the emission from the tail region of the Boomerang nebula as described by~\citet{Xin2019}, with a uniform disk with spatial width 0.25\degree\ centered around (R.A., decl.) = (336\degree.68, 60\degree.88) and power-law spectrum. Figure \ref{fig:fermi_spectrum} shows the spectral energy distribution (SED) of 4FGL J2229.0+6114. We used the same spectral model as the one reported in the 4FGL-DR2 catalog (\textit{PLSuperExpCutoff2}). The SED was extracted using the {\tt gta.sed()}
method, which fits the flux normalization of the source in each energy bin, using a power-law with a fixed index of -2. We see no emission above 50~GeV (TS of the spectral points is below a threshold value of 4); the differential upper limit in the energy range 50.7~GeV to 2~TeV is $\mathrm{2.91\times10^{-7}\ MeV\ cm^{-2}\ s^{-1}}$ at 95\% confidence level (CL). As outlined in~\citep{Abdo_2013}, pulsar spectra in the LAT energy range should cut off exponentially around a few GeV; for this reason, we conservatively only considered the measurements above 50~GeV as PWN emission in order to cut most of the pulsed emission from 4FGL J2229.0+6114. 

\subsection{VERITAS observation and analysis}
\label{sec:VERITAS}
\begin{figure}[h!]
\begin{center} 
  \includegraphics[width=0.45\textwidth]{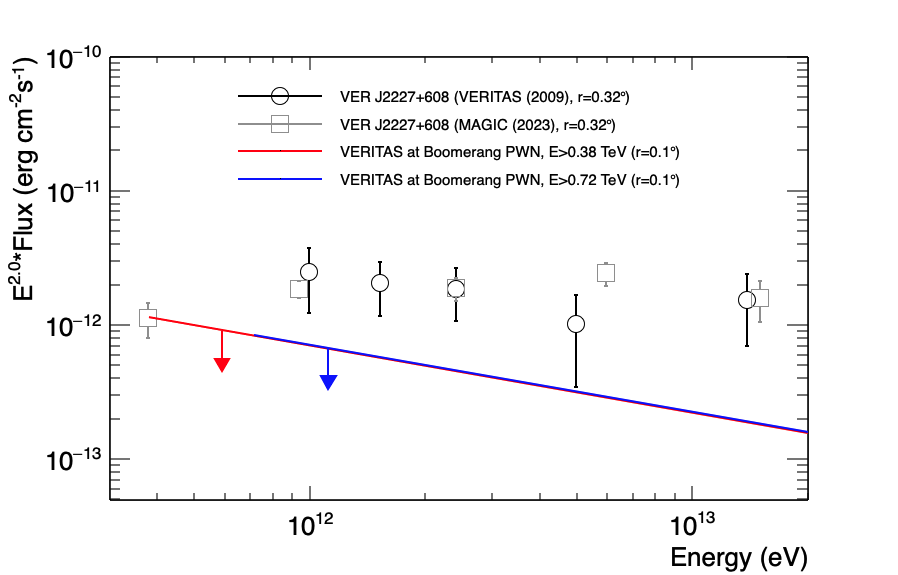}
  \caption{Upper limits of VERITAS at the location of Boomerang PWN with two event selections. VHE gamma-ray emission near the tail region of SNR~106.3+2.7 measured by the IACTs, VERITAS~\citep{Acciari2009} and MAGIC~\citep{2023A&A...671A..12M}, is shown as comparisons.}
  \label{fig:veritas_spectrum}
\end{center}
\end{figure}

\begin{deluxetable*}{ccc}[]
\label{tab:VERITAS_UL}
\tablecaption{VERITAS upper limits with 99\% confidence level}
\tablecolumns{3}
\tablehead{ E$_{threshold}$ (TeV)   & index & Upper limits (10$^{-12}$~s$^{-1}$~cm$^{-2}$) } 
\startdata   
0.38 & 2.0 & 1.10 \\
0.38 & 2.5 & 1.23 \\
0.35 & 3.0 & 1.52 \\
0.79 & 2.0 & 0.40 \\
0.72 & 2.5 & 0.48 \\
0.72 & 3.0 & 0.49 \\
\enddata
\end{deluxetable*}
VERITAS is an array of four imaging atmospheric Cherenkov telescopes (IACTs), located near Tucson, Arizona, designed to measure gamma rays of energies from 100~GeV up to $>$ 30~TeV. Each telescope has a field of view of 3.5\degree, and the array can detect a point-like source with 1\% of the Crab PWN flux at 5$\sigma$ significance within 25 hours. VERITAS has an angular resolution of $\sim$0.1\degree~\citep{Park2015}. 
The VERITAS Collaboration previously reported the detection of gamma-ray emission from the region of SNR~G106.3+2.7 with 33.4 hours~\citep{Acciari2009} based on data collected in the 2008 epoch. The TeV emission was observed near the center extended radio emission (see \ref{fig:SNRreg}) rather than the location of the Boomerang PWN. From 2009 to 2010, VERITAS accumulated an additional 22.3 hours with a changed array configuration where one telescope was moved to make the array more symmetric, which improved the sensitivity of the array~\citep{Perkins2009}. Combined with the previous data set, we used a total exposure of 57.7 hours for the analysis at the location of the Boomerang PWN. As the extension of the PWN measured in radio and X-ray is smaller than the angular resolution of VERITAS, the analysis was performed with an assumption that the emission is a point-like source. Standard VERITAS analysis was performed with two independent analysis methods. Two different event selections were used for the analysis: one selection was optimized to search for emission that was 2-10 \% of the Crab Nebula strength and one selection was optimized to search for emission weaker than 2\% of the Crab Nebula strength. The event selections optimized to search for the weaker emission reject the largest fraction of background events, resulting in a higher energy threshold.
No strong gamma ray emission was detected at the location of the Boomerang PWN. Upper limits at the 99\% confidence level for two different energy thresholds were calculated based on the assumption of a power-law spectral energy distribution with a spectral index ranging from 2 to 3. The results are shown in Table~\ref{tab:VERITAS_UL}. These upper limits are shown in the Figure~\ref{fig:veritas_spectrum} together with the SED of known VHE gamma ray emission in the tail region of SNR~G106.3+2.7. 
\section{Broadband spectral energy distribution}
\label{sec:SED}
In this section, fer modelling of the multi-wavelength data from the radio to TeV band. We present analyses using three different leptonic SED models, all one-zone (homogeneous) models with varying degrees of spectral detail. As there was no detection of gamma-ray emission above 3~GeV from the Boomerang PWN, we will not consider the effect of adding hadronic components to the model in this paper. We will include hadronic components in a future paper reviewing the SEDs from the entire SNR. The morphological complexity of the Boomerang PWN and larger-scale system is such that all three models used in this paper are highly simplified. And as can be seen in Figures \ref{fig:ChandraImage} and \ref{fig:NuSTARImage}, the X-ray centroid in both the soft and hard bands is offset from the peak in the radio band, suggesting the possibility that Boomerang is a multi-zone system. By using these one-zone models we assume that both the X-ray and radio emission originate from the same source and physical processes as part of a single system. This assumption has been made for other PWNe with similar offsets between X-ray and radio peak emission, such as the PWNe associated with SNRs G54.1$+$0.3, G327.1$+$1.1 and MSH 15$-$56 \citep{Lang2010, Temim2015, Temim2013}. Despite the simplification made in using these aforementioned models, we hope to obtain general constraints on important parameters which could guide more complex models specialized to the unique characteristics of the Boomerang system. For the radio band, we adopted the flux data from \citet{Kothes2006}. We used the \chandra\ and \nustar\ X-ray spectra after correcting for ISM absorption. Since we applied one-zone SED models, both radio and X-ray spectra were extracted from the same $r=100$\asec\ region around the pulsar position. The \fermi\ flux upper limits were derived from the analyses described in \S2.2. As the difference between the \veritas\ 0.38--30 TeV and 0.72-30 TeV upper limits found in \S2.3 is small, we chose to use the former value. Note that the  source extraction regions for \fermi\ and \veritas\ analyses are subject to the telescope PSF sizes which are larger than $r=100$\asec. Throughout the SED modelling, we consider two contrasting source distances of 0.8 and 7.5 kpc suggested by the H I velocity \citep{Kothes2001} and the pulsar's DM measurement \citep{Abdo2009b}, respectively. 

Below we present three different SED models. \texttt{NAIMA} models radiative SEDs for a given time-independent electron energy distribution (\S\ref{sec:NAIMA}). Although this is a simplistic approach, we attempt to constrain several PWN parameters such as B-field and electron spectral index. In \S\ref{sec:GAMERA}, we applied the time-dependent SED model package \texttt{GAMERA} \citep{Hahn2016} to the multi-wavelength SED data. This time-dependent approach, which assumes PWN free expansion, revealed several challenges in fitting the PWN SED data and size simultaneously, and further constrained multiple PWN properties. However, both \texttt{NAIMA} and \texttt{GAMERA} proved to be overly simplistic. As can be seen from visual inspection of the radio and X-ray flux data, a power-law fit to the X-ray data will undershoot the observed radio data, a phenomena seen for only a few other PWNe (see \citet{Hattori2020} for example). The best fit results from both \texttt{NAIMA} and \texttt{GAMERA} confirm the difficulty in reproducing both the radio and X-ray data with such simplistic models. We therefore also considered the PWN evolution using the more complex, dynamical SED model developed by \citet{Gelfand2009} in \S\ref{sec:DynamicSED}. The model has been widely used for modelling SED data of various PWNe including the Crab nebula, G21.5$-$0.9 and  composite SNR-PWN systems (e.g., \citet{Hattori2020}). The dynamical PWN model allowed us to track a full evolution path from the free expansion, SNR reverse shock interaction, and re-expansion phases. Both the SED and PWN radius are modeled as a function of time in comparison with the observation data. In this physically motivated approach, we determine the current B-field, pulsar's true age, expansion velocity, and its current evolution stage.

\subsection{NAIMA SED model}
\label{sec:NAIMA}
In order to estimate initial PWN parameters from the SED data, we relied on the \texttt{NAIMA V0.10.0} python package \citep{Zabalza2015}. \texttt{NAIMA} is a  time-independent, one-zone SED model used to generate multiple radiative model components from an assumed particle energy distribution. We fit the multi-wavelength SED data assuming that the electron distribution is in the form of a power-law model $A(E_e/E_{0})^{-p}$ between $E_e=E_{\rm min}$ and $E_{\rm max}$. While a hard cutoff at $E_e=E_{\rm min}$ is not physically motivated, it is implemented to simplify the model. We generated leptonic radiation models with synchrotron radiation, ICS and synchrotron self Compton (SSC) components. We adopted the cosmic microwave background (CMB) as a seed photon source for the ICS component. No IR seed photons were added so as to maintain the ICS component of the model as a lower limit. Furthermore, including a variable IR seed photon component would only add to the model's degeneracies, an already significant issue which we discuss below. We consider an IR photon field in the more elaborate dynamical model in Section \ref{sec:DynamicSED}. While the initial physical parameter estimates of the emitted plasma provided by \texttt{NAIMA} are useful, the model does not directly provide any insight as to where this emitting plasma came from, and -- due to the evolution of the PWN inside the SNR, especially once it collides with the reverse shock -- the particle spectrum is unlikely to be well described by a single or broken power-law. Our fitting results confirm \texttt{NAIMA's} inadequacy in describing the Boomerang PWN's particle spectrum.
\begin{figure*}[t!]
    \centering
     {{\includegraphics[width=0.44\textwidth]{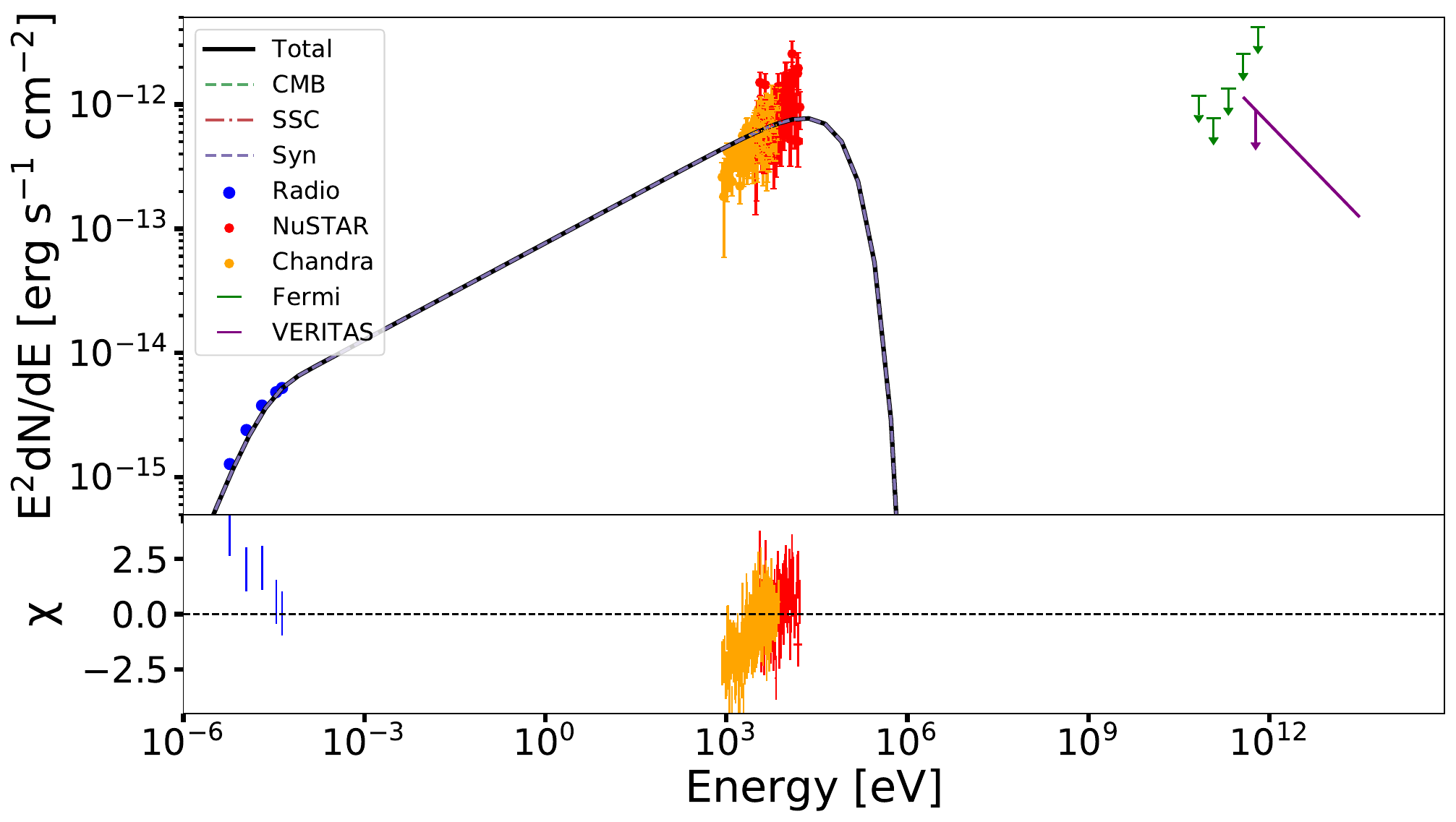} }} 
    \qquad 
    \includegraphics[width=0.44\textwidth]{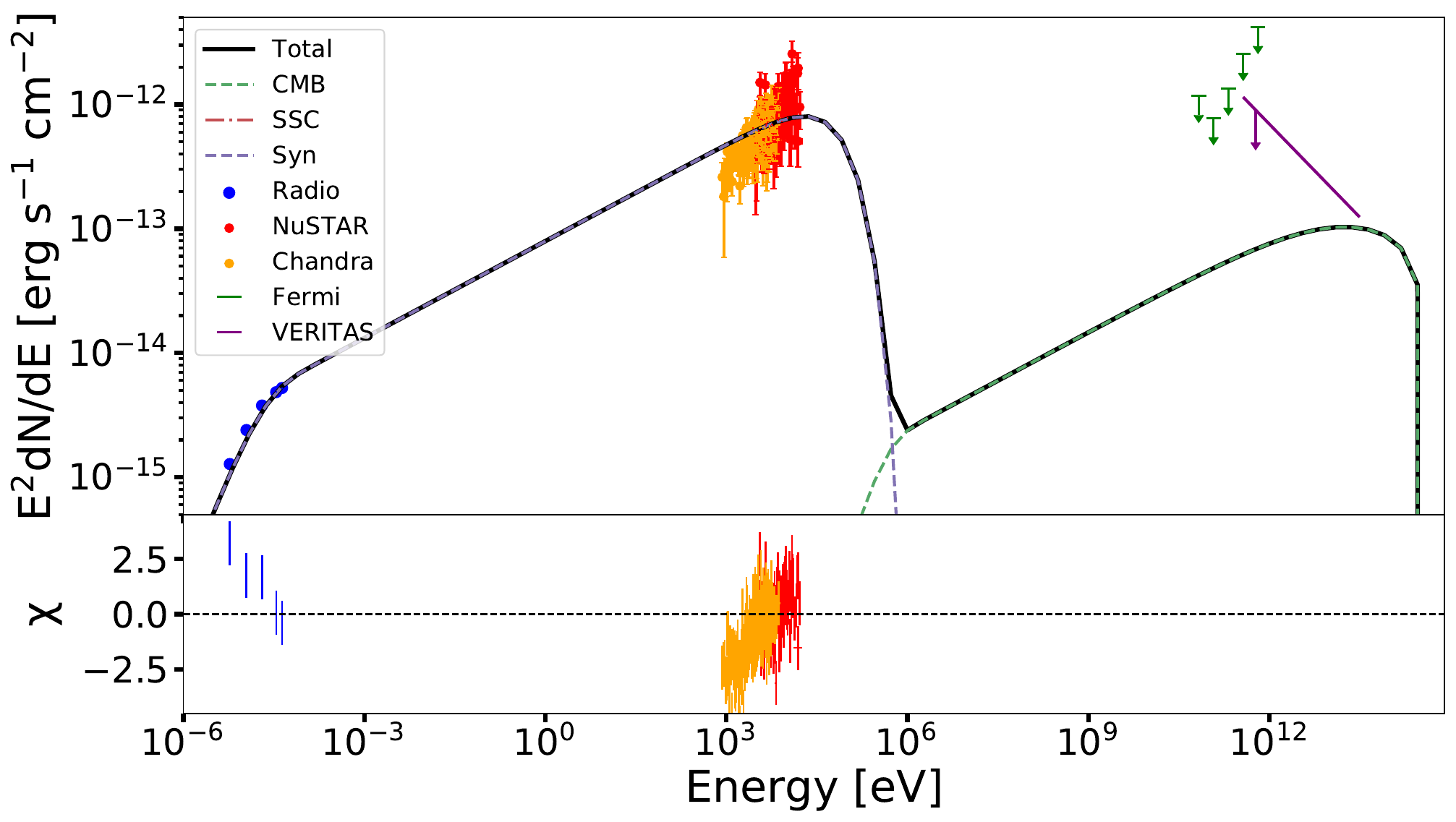} 
    \caption{NAIMA simple power-law SED fits with B-field set to 2.6 mG (left) and $5~\mu$G (right). The complete set of parameters for each fit, as well as for other B-field scenarios, is shown in Table \ref{tab:NAIMAparams}.
    }
    \label{fig:NAIMAplots}
\end{figure*}
\begin{deluxetable*}{lccccccccc}[t!]
\tablecaption{Fit parameters from various leptonic NAIMA model cases.}
\tablecolumns{8}
\tablehead{ \colhead{Distance [kpc]} & \colhead{$B$ [$\mu$G]}  & 
\colhead{$p$}  & \colhead{$E_{\rm min}$ [GeV]} & \colhead{$E_{\rm max}$ [TeV]} & \colhead{Electron energy [erg]} & \colhead{Magnetic field energy [erg]\tablenotemark{a}} &
\colhead{$\eta_E$} & \colhead{$\eta_B$}}
\startdata   
0.8 & 2600 & 2.5 & 0.32 & 22 & $2.3 \times\ 10^{41}$ &  $2.0 \times\ 10^{48}$ & $\sim0$ & $\sim1$ \\
0.8 & 100 & 2.5 & 1.64 & 120 & $2.9 \times\ 10^{43}$ & $2.9 \times\ 10^{45}$ & 0.01 & 0.99 \\
0.8 & 10 & 2.5 & 5.18 & 400 & $1.0 \times\ 10^{45}$ & $2.9 \times\ 10^{43}$ &  0.97 & 0.03 \\
0.8 & 5 & 2.5 & 7.33 & 500 & $2.8 \times\ 10^{45}$ &  $7.3 \times\ 10^{42}$ & 0.998 & 0.002 \\
\hline
7.5 & 2600 & 2.5 & 0.32 & 22 & $2.0 \times\ 10^{43}$ &  $1.6 \times\ 10^{51}$ & $\sim0$ & $\sim1$ \\
7.5 & 100 & 2.5 & 1.64 & 120 & $2.5 \times\ 10^{45}$ & $2.9 \times\ 10^{48}$ & 0.001 & 0.999 \\
7.5 & 10 & 2.5 & 5.18 & 400 & $9.1 \times\ 10^{46}$ & $2.9 \times\ 10^{46}$ &  0.76 & 0.24 \\
7.5 & 5 & 2.5 & 7.33 & 500 & $2.5 \times\ 10^{47}$ &  $7.3 \times\ 10^{45}$ & 0.97 & 0.03 \\
\enddata
\tablenotetext{a}{$\eta_E$ and $\eta_B$ are calculated by dividing the electron and magnetic field energy by their sum, respectively.}
\label{tab:NAIMAparams}
\end{deluxetable*}

We first focused on reproducing the radio spectral break at $\sim 5$ GHz and X-ray spectra. 
We found that the radio data are adequately fit by various sets of $B$ and $E_{\rm min}$ values, as listed in Table \ref{tab:NAIMAparams}. We have shown that the radio spectral break can be caused by the selected $E_{\rm min}$. The origin of the break in the electron spectrum, used here to fit the radio break, cannot be determined with the model. Given the degeneracy between $B$ and $E_{\rm min}$, we did not find $B=2.6$ mG as a unique solution to reproduce the radio data, contrary to the results of \citet{Kothes2006}. 
On the other hand, the electron spectral index was constrained to $p = 2.5$ by fitting the radio and X-ray spectra with a synchrotron model component. We expect that $(p - 1)/2 = \Gamma_{x} - 1$. We therefore infer from $p = 2.5$ that $\Gamma_{x} = 1.75$, significantly larger than any of the X-ray photon indices found in Table \ref{tab:xspecfit}, confirming that \texttt{NAIMA} cannot adequately fit both the radio and X-ray data. However, the parameters serve as an initial estimate. For each assumed source distance (0.8 \& 7.5 kpc), a list of the model parameters for four representative cases (including $B = 2.6$ mG) is shown in Table \ref{tab:NAIMAparams} and two representative SEDs are plotted in Figure \ref{fig:NAIMAplots}. In the gamma-ray band, the ICS and SSC components remain below the \fermi\ and \veritas\ flux upper limits as long as B $>$ 5~$\mu$G. 
This is because, as the magnetic field is decreased, a larger number of electrons is required to fit the radio and X-ray data and thus enhances the model GeV--TeV flux through the ICS component. If we consider other seed photon sources than the CMB, the lower limit will be higher than $5~\mu$G. 

{\tt NAIMA} outputs the total energy of the electron population, $U_{E}$, integrated between $E_{\rm min}$ and $E_{\rm max}$. Assuming the B-field is uniform throughout the interior of the PWN ($r < 100\asec$), we calculated the total magnetic field energy as $U_{B} = \frac{B^{2}}{8\pi}\times V_{\rm PWN}$, where $V_{\rm PWN} =(4/3)\pi R_{\rm PWN}^3$ is the volume of the PWN. The total electron and magnetic field energies for each fit are provided in Table \ref{tab:NAIMAparams}. Given the constraints on the synchrotron radiation fluxes, increasing the magnetic field from B $\sim$1$\rm{-}10$~$ \mu$G to B $\simgt$ 100~$\mu$G requires decreasing the electron density dramatically. As a result, a fraction of the total energy allocated to PWN B-field ($\eta_{B}$) is $\sim1$ for B $\simgt$ 100$\mu$G. Such a high magnetization parameter is unusual if compared to other PWN systems -- e.g., \citet{Martin2014A} found $\eta_B = 7\times10^{-4}\rm{-}0.02$ from six PWNe including the Crab nebula (which possesses the highest $\eta_B$ value). If we attain an $\eta_{B}$ value comparable to those deduced from the other PWNe, it points to a lower B-field range of B $\simlt$ 10~$\mu$G. While these arguments as well as the suggested B-field  range (B$\sim$5$\rm{-}10$~$\mu$G) are empirical based on the time-independent one-zone SED model fitting, we will address this issue with more advanced SED models and other X-ray analysis results in the later sections. 
\newpage
\subsection{GAMERA SED model}
\label{sec:GAMERA} 
To further characterize the PWN properties, we modelled the time evolution of the particle distribution and radiation SED using \texttt{GAMERA} \citep{Hahn2016}. \texttt{GAMERA} allows us to inject leptons and track their energy distribution as they vary via radiative and adiabatic cooling over time. However, the {\tt GAMERA} modelling does not account for the interaction between the PWN and the parent SNR which makes the PWN compress and expand, altering the injected particle distribution significantly. The model assumes that the PWN is expanding with a constant velocity, and thus only tracks the recent PWN evolution. The age reflects the time since the PWN started expanding, whether that be from the time of the pulsar's birth or from the time that the reverse shock passed through the PWN. We assumed $R_{\rm PWN}$ increased linearly with time over the short lifetime of the new PWN and matched its current radius (at $t=t_{\rm PWN age}$) with the measured radio PWN size. Assuming two different sources distances (0.8 and 7.5 kpc), we adopted the corresponding PWN sizes of $r=0.4$ pc and 4 pc.
We assumed continuous particle and magnetic field injection into the PWN at the rate of the pulsar's current spin-down power $\dot{E}=2.2\times 10^{37}$ \lumcgs. A fraction of the spin-down power ($\eta_B$) is injected into the PWN in the form of magnetic potential energy $U_B$. The magnetic field B was assumed to be uniform throughout the spherical volume of the PWN. The rest of the spin-down power is injected into particle energy following a power-law distribution $A(E/E_0)^{-p}$. The normalization factor $A$ is determined by the total particle energy of $(1-\eta_g-\eta_B)\dot{E}$ where $\eta_g$ is the gamma-ray efficiency of the pulsar. We ignored the fraction of the spin down energy allocated to gamma-ray pulsations; this contribution should not significantly effect the model outputs. The injected particle distribution is evolved by three particle cooling mechanisms, i.e. adiabatic, synchrotron and ICS  components. The adiabatic cooling rate is calculated using the expansion velocity of the PWN, which we assumed to be constant ($v_{\rm PWN}=R_{\rm PWN}/t_{\rm age}$). We then calculated a radiation SED from the particle distribution at $t = t_{\rm age}$. We employed a leptonic radiative model comprised of emission from synchrotron radiation and ICS of the CMB and synchrotron-self-Compton scattering emission. Similarly to our \texttt{NAIMA} fitting, we have only considered ICS of CMB and SSC components into account to maintain the ICS component of the model as a lower limit. Still, \texttt{GAMERA} is not able to describe the data very well as summarized below. Adding other potential seed photons, such as IR, would only make fitting the data more challenging for these models, as the IC component would encroach on the \fermi\ and \veritas\ upper limits. The addition of an IR component is considered for our dynamical model in Section \ref{sec:DynamicSED}.

\begin{figure*}[t!]
\centering
        \includegraphics[width=0.43\textwidth]{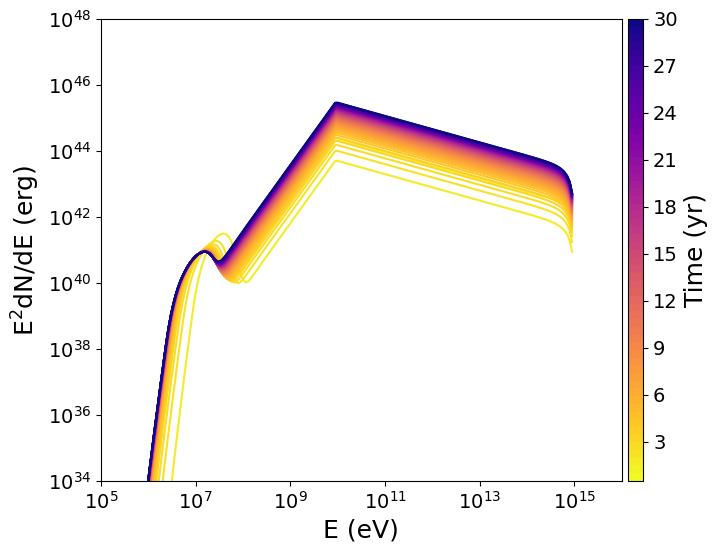}
        \includegraphics[width=0.43\textwidth]{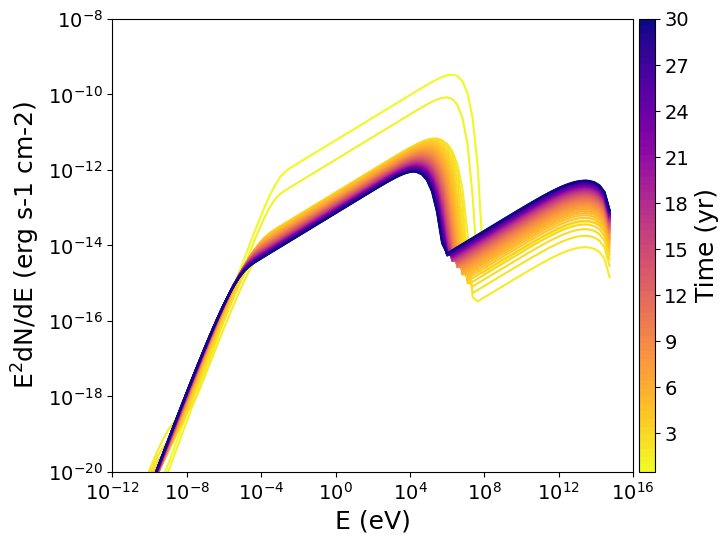}
        \includegraphics[width=0.41\textwidth]{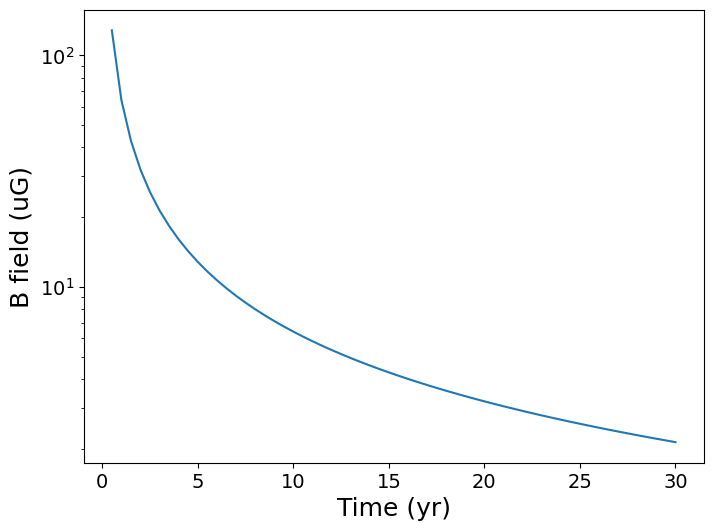}
        \includegraphics[width=0.44\textwidth]{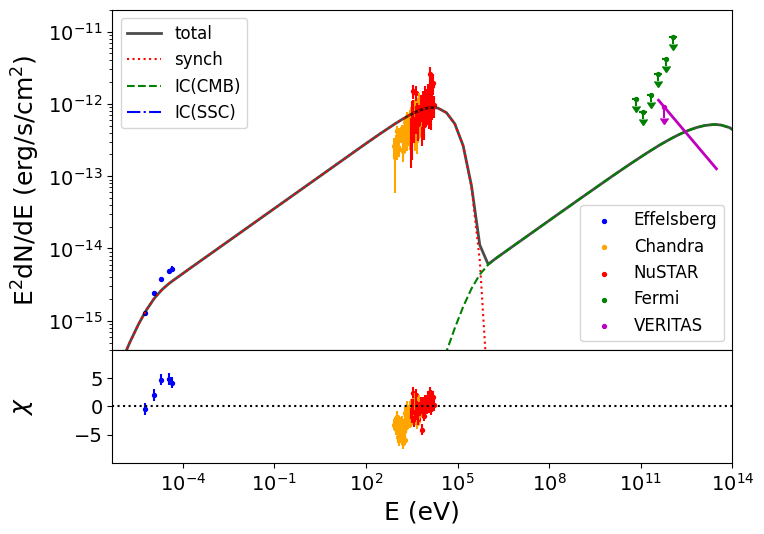}
    \caption{Time evolution of the lepton (top left) and radiation (top right) SEDs, and B-field (bottom left), and the radiation SED at the current time (bottom right) assuming $d = 0.8$ kpc.}
    \label{fig:GAMERAplots1}
\end{figure*}

We also attempted to constrain our model based on the lack of X-ray variability over the last $\sim20$ years (\S\ref{sec:spectroscopy}). Since the luminosity of synchrotron radiation is proportional to $B^2$, a sizable variation in B-field can lead to detectable X-ray flux variability. As the PWN evolves, decreasing B-field and increasing particle density cancel out with each other and result in a slower synchrotron luminosity evolution, while ICS luminosity keeps increasing as more particles are injected over time. While fitting the \texttt{GAMERA} model to the multi-wavelength SED data, we tracked the synchrotron X-ray luminosity $L_{\rm syn} (t)$ evolution and found solutions without significant X-ray variability over the last $\sim20$ years.  

\begin{figure*}[t!]
\centering
        \includegraphics[width=0.43\textwidth]{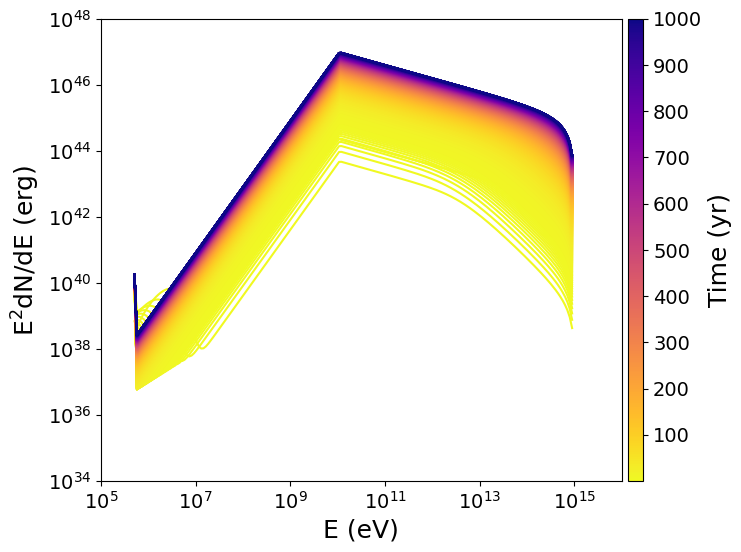}
        \includegraphics[width=0.43\textwidth]{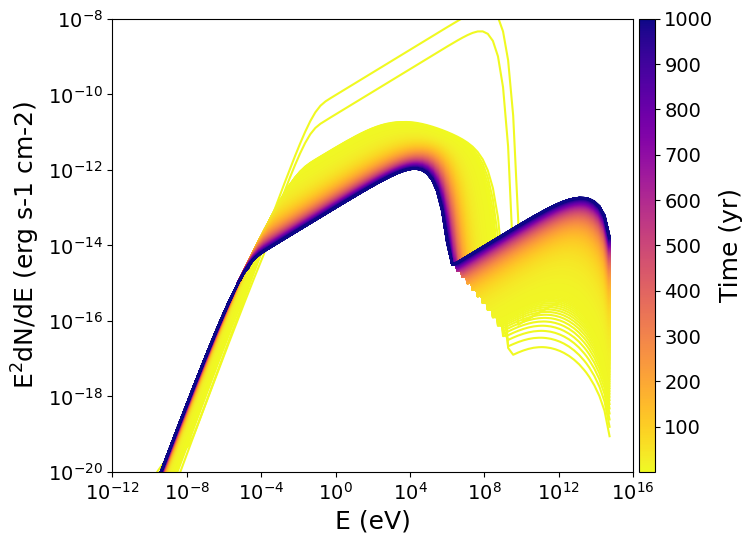}
        \includegraphics[width=0.41\textwidth]{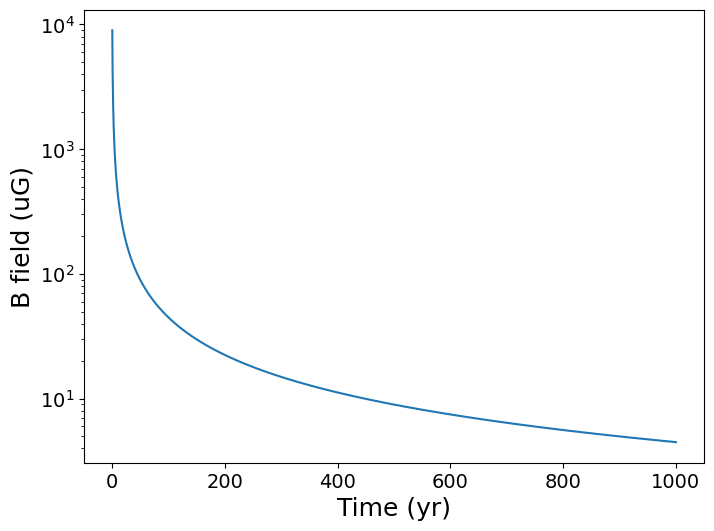}
        \includegraphics[width=0.44\textwidth]{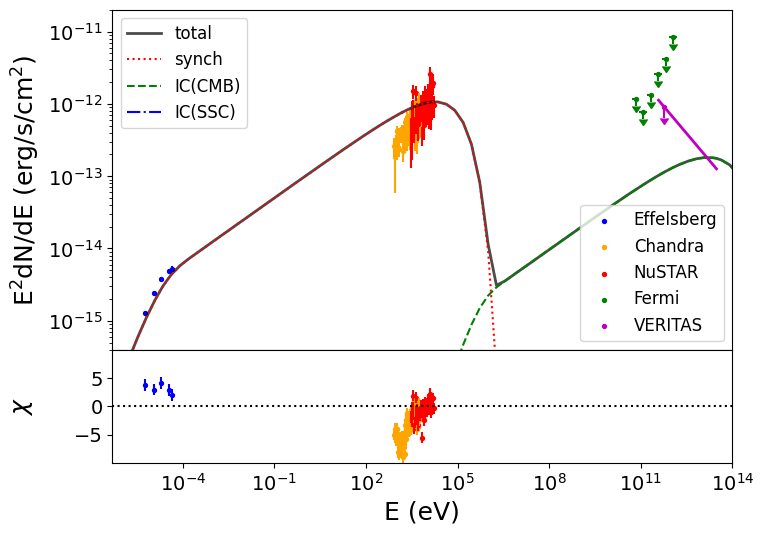}
    \caption{Same plot as Figure \ref{fig:GAMERAplots1} for $d = 7.5$ kpc. } 
    \label{fig:GAMERAplots2}
\end{figure*}
\begin{deluxetable*}{lcc}[b!]
\tablecaption{Model parameters for the case presented in \S\ref{sec:GAMERA}}
\tablecolumns{3}
\tablehead{ \colhead{Parameter}   & \colhead{$d = 0.8$ kpc} & \colhead{$d = 7.5$ kpc} } 
\startdata   
Age [yr] & 30 & 1000\\ 
$p$ & 2.4 & 2.4\\ 
$E_{\rm min}$ [GeV] & 7.9 & 10.0\\ 
$E_{\rm max}$ [PeV] & 1.0 & 1.0\\ 
$\eta_g$\tablenotemark{a} & $1\times10^{-3}$ & $7\times10^{-2}$\\
$\eta_B$ & $7\times10^{-5}$ & $1\times10^{-2}$\\ 
$B$ [$\mu$G] & 2.1 &  4.5\\ 
Electron energy [erg] & $8.9 \times 10^{45}$ & $2.9\times10^{47}$ \\ 
Magnetic field energy [erg] & $1.3 \times 10^{42}$ &  $6.3\times10^{45}$\\ 
Expansion velocity [km/s] & $1.3\times10^4$ & $3.9\times10^3$\\
\enddata
\tablenotetext{a}{The gamma-ray efficiency of PSR J2229+6114 from \citet{Abdo2009b}. $f_{\Omega}$ (Eq. (3) in the cited work) was assumed to be 1.}
\label{tab:GAMERAparams}
\end{deluxetable*}

The best-fit parameters are listed for the short and long distance cases in Table~\ref{tab:GAMERAparams}. Figure \ref{fig:GAMERAplots1} and \ref{fig:GAMERAplots2} show a time series of the particle SED, radiation SED, and magnetic field, along with the radiation SED at the current time using the best-fit parameters. Note that both GAMERA models grossly misrepresent the spectrum in the X-ray region, and have other shortcomings as discussed below.

(1) $d = 0.8$ kpc case: The lifetime of the PWN should be $t_{\rm age} \sim30$ yr for $d = 0.8$ kpc. If $t_{\rm age} \simlt 20$ yr, the model predicts that the 2002 \chandra\ observation should have detected a much higher X-ray flux than it did. If the lifetime is longer than $\tau \sim 30$ yr, too many leptons are injected into the PWN to be consistent with the GeV and TeV flux upper limits. Currently, the B-field should be as low as $B \sim 2~\mu$G (with a very low magnetization of $\eta_B = 7\times10^{-5}$). Otherwise, the model over-predicts synchrotron radiation fluxes in the radio and X-ray band. We also found that $E_{\rm min}$ is well constrained by the radio spectral break and the overall flux normalization of electrons (which depends sensitively on $E_{\rm min}$ for a given $\dot{E}$ and lifetime). The expansion velocity should be high ($V_{\rm PWN} = 1.9\times10^4$ km/s) in order to reach the observed PWN radius within $\sim30$ yr. To relax these stringent constraints, we introduced an extra, non-radiative energy loss term in {\tt GAMERA} via escaping leptons. Only if we assumed a short particle escaping time ($t_{\rm esc} < 0.05$ yr), were we able to fit the SED data with higher $\eta_B$ values and older ages. However, such a short particle escape time is unrealistic as it is shorter than the light crossing time of the PWN. 

(2) $d = 7.5$ kpc case:  In contrast, the larger distance allows the PWN to radiate away its rotational energy over a more extensive period of time. For example, over 1,000 yr significantly more particle energy can be injected, totalling $3.1\times10^{47}$ ergs. While the injected electron spectral parameters such as $p$, $E_{\rm min}$ and $E_{\rm max}$ are similar between the two distance cases (Table~\ref{tab:GAMERAparams}), $\eta_B$ is higher at $1\times10^{-2}$ while the current B-field is $4.5~\mu$G. The expansion velocity is $3.9\times10^3$ km/s, which is more in line with the observed PWN velocities ($v_{\rm PWN} \sim 1,000\rm{-}1,500$ km/s) from the Crab nebula and Kes-75 \citep{Bietenholz1991, Reynolds2018}.   

Overall, we found that the larger distance allows for an older age (i.e. longer particle injection time) and higher magnetization parameter. 
In either case, the current B-field needs to be as low as $B\sim5~\mu$G in order to match the low radiation efficiency ($L_{\rm syn}/\dot{E}$). Consequently, the radio spectral break cannot be caused by synchrotron cooling at $B = 2.6$ mG as suggested by \citet{Kothes2006}. Alternatively, we found that the radio break energy is directly related to the minimum energy ($E_{\rm min}$) in the injected particle energy distribution. Unlike the {\tt NAIMA} model, the added complexities of the {\tt GAMERA} model broke the degeneracy between $B$ and $E_{\rm min}$. In both cases, $E_{\rm min} \sim 10$ GeV fit the radio SED data well, and thus we attribute the break to the PWN's intrinsic particle injection distribution. 

\subsection{Dynamical PWN evolution model}
\label{sec:DynamicSED}

In this section, we explore the time evolution of the Boomerang PWN using the dynamical PWN model \citep{Gelfand2009}.  
The SED model takes input parameters for the PWN, SNR and its environment. This model evolves a homogeneous spherical bubble of relativistic electrons and magnetic field, injected according to the pulsar spin-down luminosity and its evolution, following the dynamics of its expansion into first the expanding ejecta of a spherical SNR, and including the eventual compression by the returning reverse shock and subsequent expansion into the interior of a Sedov blast wave. More details on the model description and applications to other PWNe can be found in \citet{Gelfand2009, Gelfand2017, Hattori2020, Burgess2022}. The physically motivated model tracks the time evolution  of particle energy distribution, radiative SED and PWN properties (e.g., $B$ and $R_{\rm PWN}$) by considering particle injection and energy loss due to radiative and adiabatic cooling at each time step. The size and bulk velocity of the PWN are calculated by the pressure balance between the pulsar wind and SNR ejecta. Setting up each model run begins with determining the pulsar's properties. Given the observed current spin-down power $\dot{E}$  ($=2.2\times10^{37}$ \lumcgs) and characteristic age $t_{\rm ch}$  (= 11 kyr), we first derive the system's true age ($t_{\rm age}$) and  initial spin-down luminosity $\dot{E}_0$  
\newpage
\begin{eqnarray}
    t_{\rm age} & = & \frac{2t_{\rm ch}}{p-1} - \tau_{\rm sd} \\
    \dot{E}_0 & = & \dot{E} \left(1 + \frac{t_{\rm age}}{\tau_{\rm sd}} \right)^{\frac{p+1}{p-1}}, 
\end{eqnarray}
where $p$ and $\tau_{\rm sd}$ are the pulsar's braking index and spin-down timescale, respectively. These input parameters fully characterize the pulsar as a particle injection source. 

In the model, the pulsar injects leptons and magnetic field at each time step by partitioning the time-dependent spin-down power $\dot{E}(t)$ into $(1-\eta_B) \dot E(t)$ and $\eta_B \dot E(t)$, respectively. The allocated electron energy is distributed between $E_e = E_{\rm min}$ and $E_{\rm max}$ following a broken power-law model. The evolution of both SNR forward and reverse shocks is separately calculated by going through the free expansion and Sedov-Taylor phases. The density profile of SNR ejecta, which follows  $\rho_{\rm ej} (r) \propto r^{-9}$ until reaching the density of the ISM, is used to calculate the pressure balance between the pulsar wind and ejecta. The ISM density effects the timescale of the SNR evolution and SNR reverse shock. After the SNR reverse shock hits the PWN, the ISM density plays an important role as it imposes additional pressure on the pulsar wind. These pressure factors determine the size and bulk velocity of the PWN at each time step. The injected leptons lose their energy via radiative and adiabatic cooling as the PWN expands over time. The radiative cooling in the model takes into account synchrotron and ICS components which are provided as radiative SED model output. At some point, the SNR reverse shock reaches the PWN and compresses it to the point where the pulsar wind and reverse shock are in a pressure equilibrium before the PWN begins expanding again. 

\begin{figure}[t!]
\begin{center}
    \centering
    \includegraphics[width=0.45\textwidth]{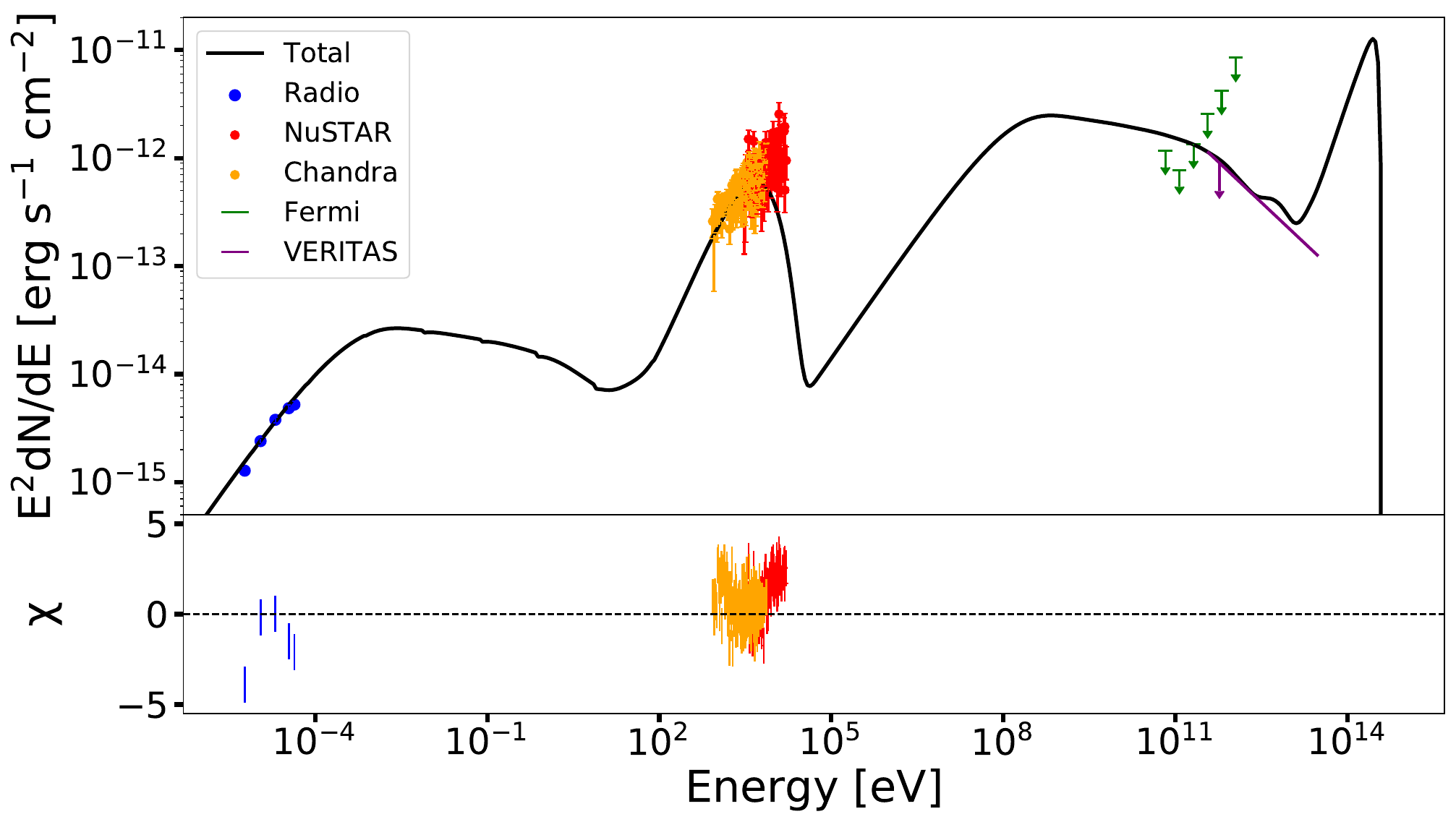} 
    \qquad 
    {{\includegraphics[width=0.45\textwidth]{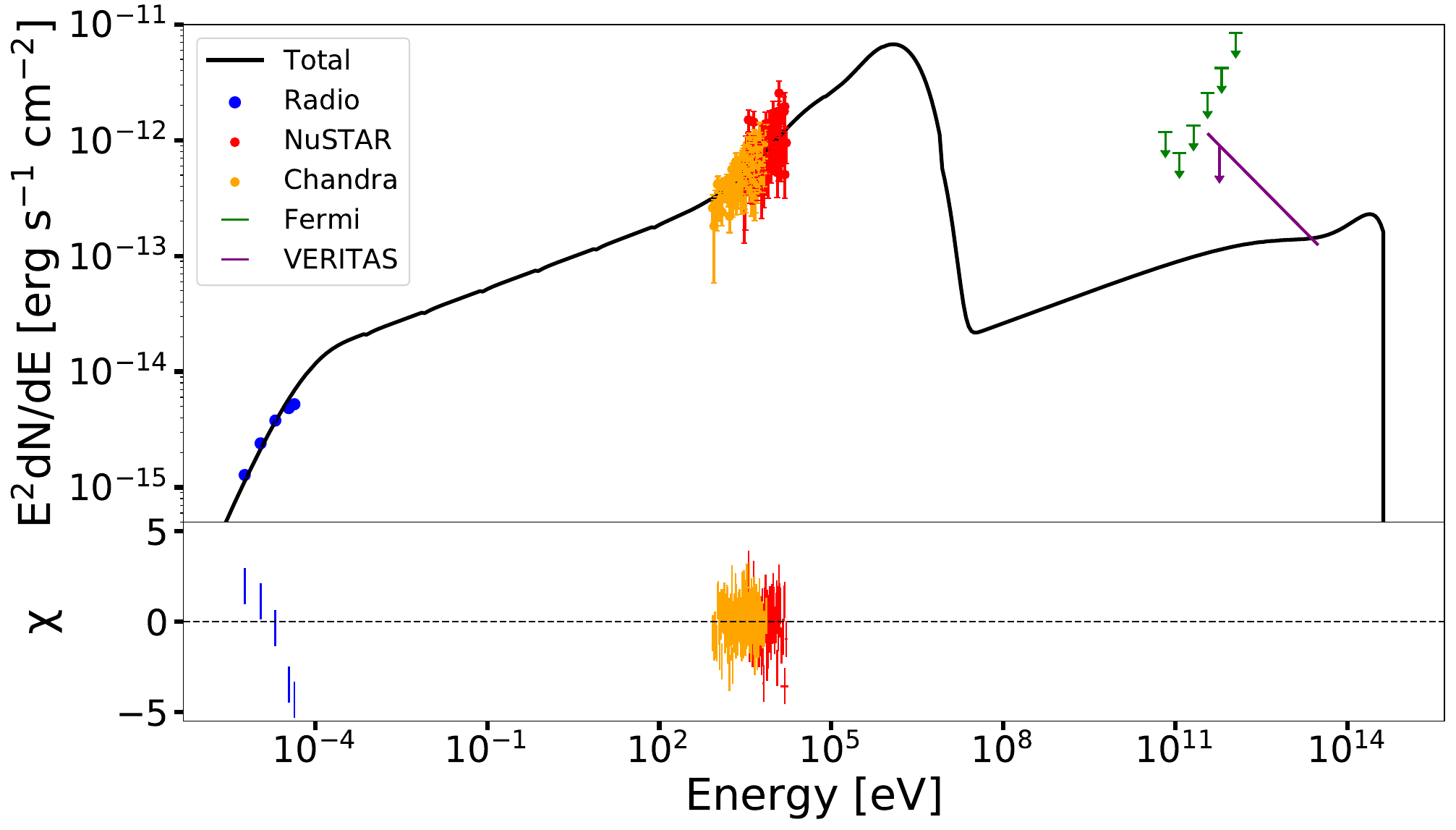} }} 
    \includegraphics[width=0.46\textwidth]{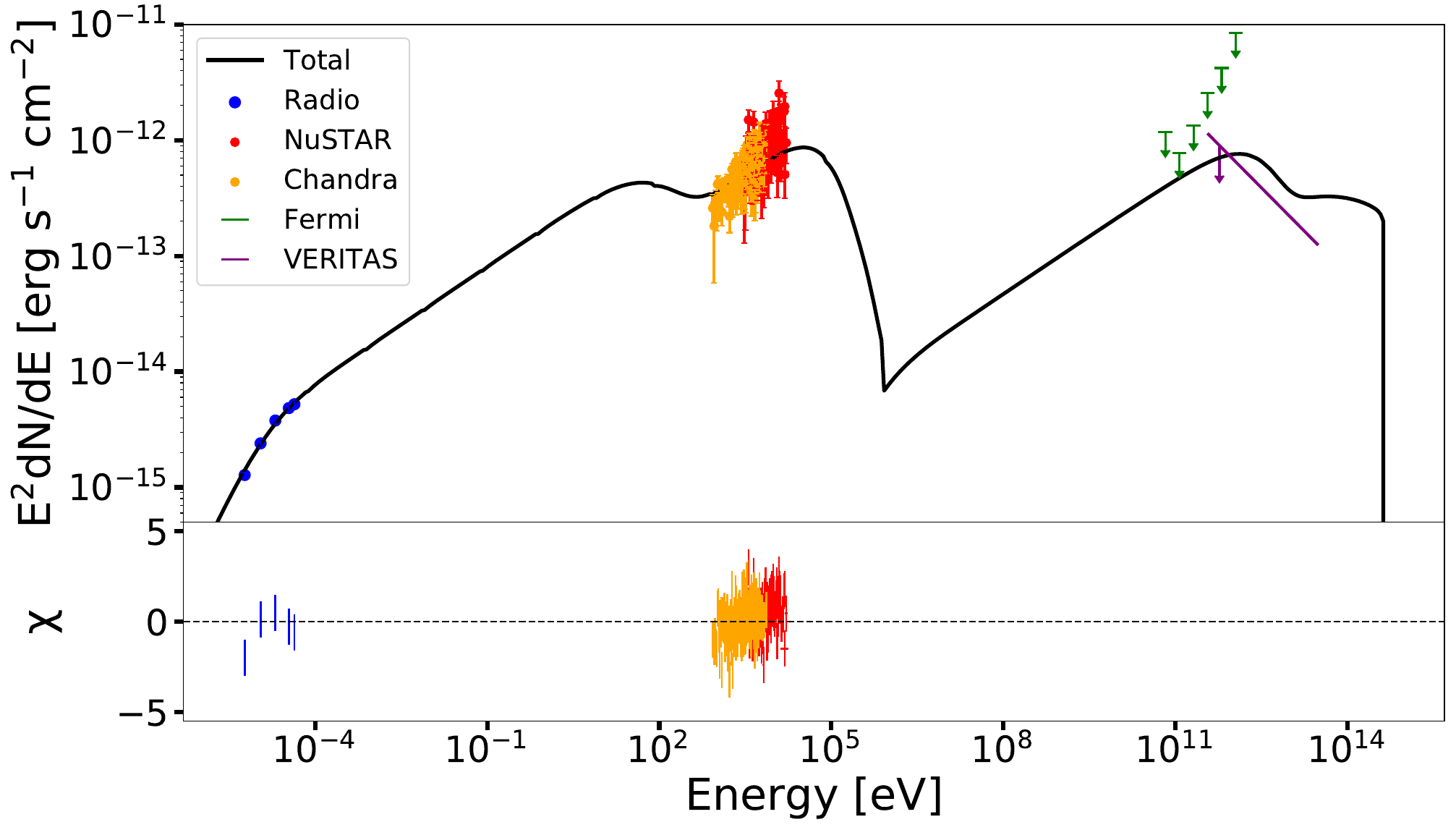} 
    \caption{Top left panel: PWN evolution SED model for case A assuming $d=0.8$ kpc. Top right panel: Case B for $d = 7.5$ kpc. Bottom panel: Case C for $d = 7.5$ kpc. The model parameters for these cases can be found in Table 4. 
    }
    \label{fig:sed_evolution}
\end{center}
\end{figure}

We aimed to reproduce both the multi-wavelength SED data and PWN size with the dynamical model. We note that the flux upper limit obtained by \veritas\ changes depending on the assumed spectral index (see Table \ref{tab:VERITAS_UL}). For the dynamical SED fitting, we used the \veritas\ upper limit at the decorrelation energy because the sensitivity of the limit to changes in the spectral index is lowest at this energy. As shown in Figure \ref{fig:sed_evolution}, we considered the upper limit optimized for $2\%$ of the Crab Nebula strength for the model fitting, where the decorrelation energy is 1.12 TeV. As we did for the \texttt{NAIMA} and \texttt{GAMERA} model fitting, we considered the case for both $d = 0.8$ and 7.5 kpc below. We initially only use CMB as a seed photon source for the ICS component; we then test the effect of adding an IR field to the model. We recognize that the detailed radio and X-ray structure of the PWN (pulsar with small X-ray nebula in the interior of the boomerang-shaped radio arc) is not well represented by a homogeneous sphere, but as with the two previous models, we hope to obtain some general insight into the possible nature and evolution of the PWN with this tool. 

(1) $d = 0.8$ kpc case: Given the very low radiation efficiency ($L_{\rm radio} / \dot{E} = 3\times10^{-8}$ and $L_{\rm X} / \dot{E} = 7\times 10^{-6}$), it is extremely difficult to allocate the pulsar's expended rotational energy without overshooting the flux data.To suppress the synchrotron radiation in the radio and X-ray bands, we need to minimize the number of injected leptons and PWN B-field. We were able to satisfy both the low radiation efficiency and  compact PWN size ($r=0.4$ pc at $d = 0.8$ kpc) with the following (Case A) SED model. For $d = 0.8$ kpc, any free-expansion phase solution that matched the flux data largely overestimated the PWN size. We therefore had to consider a re-expanding PWN after its interaction with the SNR reverse shock. We were able to roughly reproduce all the radio, X-ray and gamma-ray fluxes by allocating the particle energy to the unobserved energy bands such as MeV and $>100$ TeV energies. We arrived at a very low $\eta_B$ and high ISM density. The predicted PWN radius ($r = 0.35$ pc) is roughly consistent with the Boomerang PWN size. However, the predicted X-ray spectral shape is not consistent with the \nustar\ data and also requires an extremely high braking index ($p = 5.6$) or very young pulsar age ($t_{\rm age} = 640$ yr). While the Case A model almost works for reproducing the observation data, we do not consider it compelling due to the unreasonable parameter values. For example, the SNR's interaction with the high ISM density in Case A (160 cm$^{3}$) at a distance of 0.8 kpc and age of 640 years would produce the brightest thermal SNR ever observed by a large factor. Under the assumption that the PWN confines all of the leptons injected over its entire lifetime, it is nearly impossible to model the observed PWN's faint emission if the source distance is 0.8 kpc.
\begin{deluxetable*}{lcccc}[t!]
\tablecaption{Model parameters for the four cases presented in \S\ref{sec:DynamicSED}} 
\tablehead{\colhead{Model parameter} & \colhead{Case A} & \colhead{Case B} & \colhead{Case B'} & \colhead{Case C}}
\startdata 
Source distance [kpc] & 0.8 & 7.5 & 7.5 & 7.5 \\ 
SN explosion energy [ergs] & $3.0 \times 10^{50}$ & $3.6 \times 10^{51}$ & $4.0 \times 10^{51}$ & $1.9 \times 10^{51}$ \\
SN ejecta mass [$M_\odot$] & 1.6 & 4.2 & 4.9 & 2.0 \\
ISM density [cm$^{-3}$] & 160 & 0.3 & 0.3 & 0.9 \\
Magnetic field strength [$\mu$G] & 1.5 & 3.7 & 2.9 & 2.5 \\
PWN radius [pc] & 0.35 & 3.8 & 4.6 & 2.9 \\
Pulsar braking index & 5.6 & 2.9 & 2.8 & 3.1 \\
Pulsar spin-down timescale [kyr] & 3.9 & 8.9 & 9.1 & 7.7 \\
True age [kyr] & 0.64 & 1.8 & 2.2 & 2.1 \\
Initial spin-down power [ergs/s] & $2.7\times 10^{37}$ & $2.2 \times 10^{37}$ & $3.4 \times 10^{37}$ & $3.5 \times 10^{37}$ \\
Wind magnetization ($\eta_B$) & $8\times10^{-8}$ & 0.006 & 0.005 & 0.0007 \\
$E_{\rm min}$ [GeV] & 0.4 & 22.3 & 20.0 & 8.7 \\
$E_{\rm max}$ [PeV] & 0.2 & 4.0 & 3.5 & 1.2 \\
$E_{\rm break}$ [TeV] & 0.14 & 335.77 & 300.00 & 0.01 \\
p$_{1}$ & 1.5 & 2.6 & 2.6 & 1.6 \\
p$_{2}$ & 3.1 & 1.3 & 1.3 & 2.3
\enddata 
\centering 
\tablenotetext{}{The SED plots for Case A, Case B, and Case C are shown in Figure \ref{fig:sed_evolution}. The parameters p$_{1}$ and p$_{2}$ are the particle indices below and above $E_{\rm break}$, respectively.} 
\label{tab:evolutionary_table}
\end{deluxetable*}
\begin{figure*}[t!]
\centering
        \includegraphics[width=0.49\textwidth]{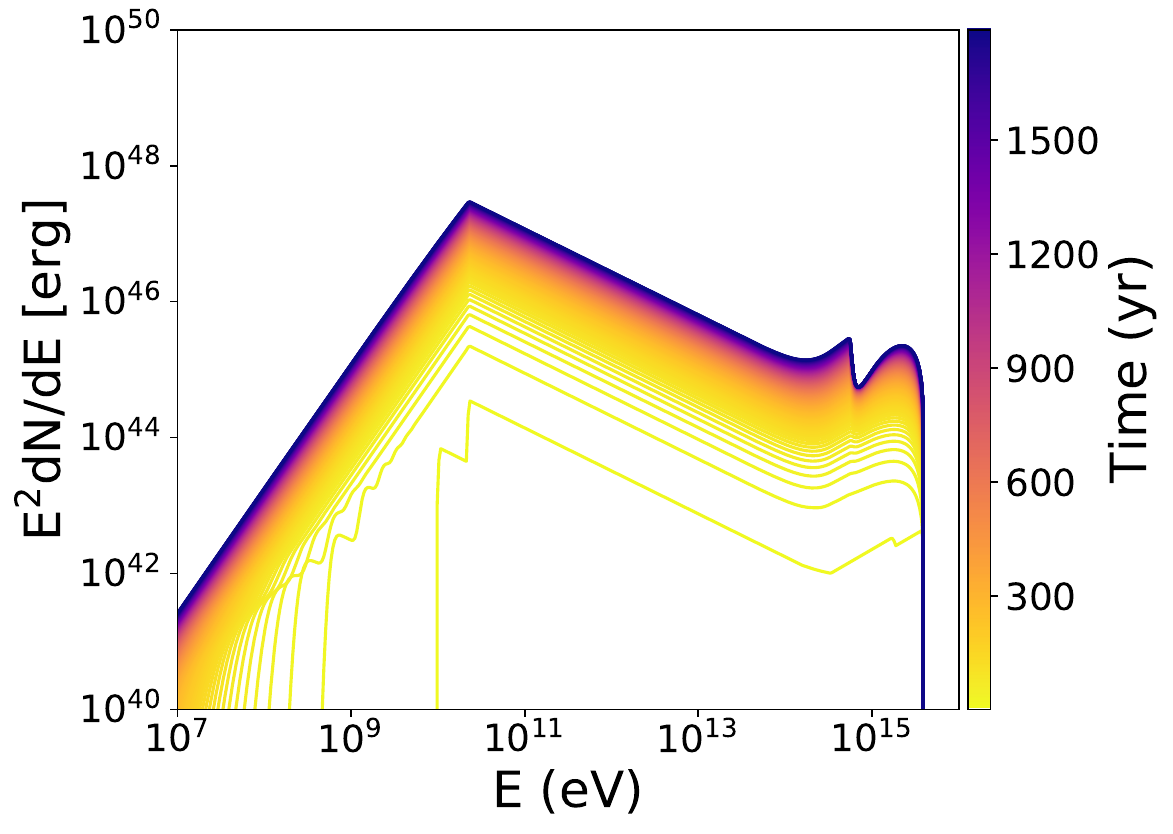}
        \includegraphics[width=0.49\textwidth]{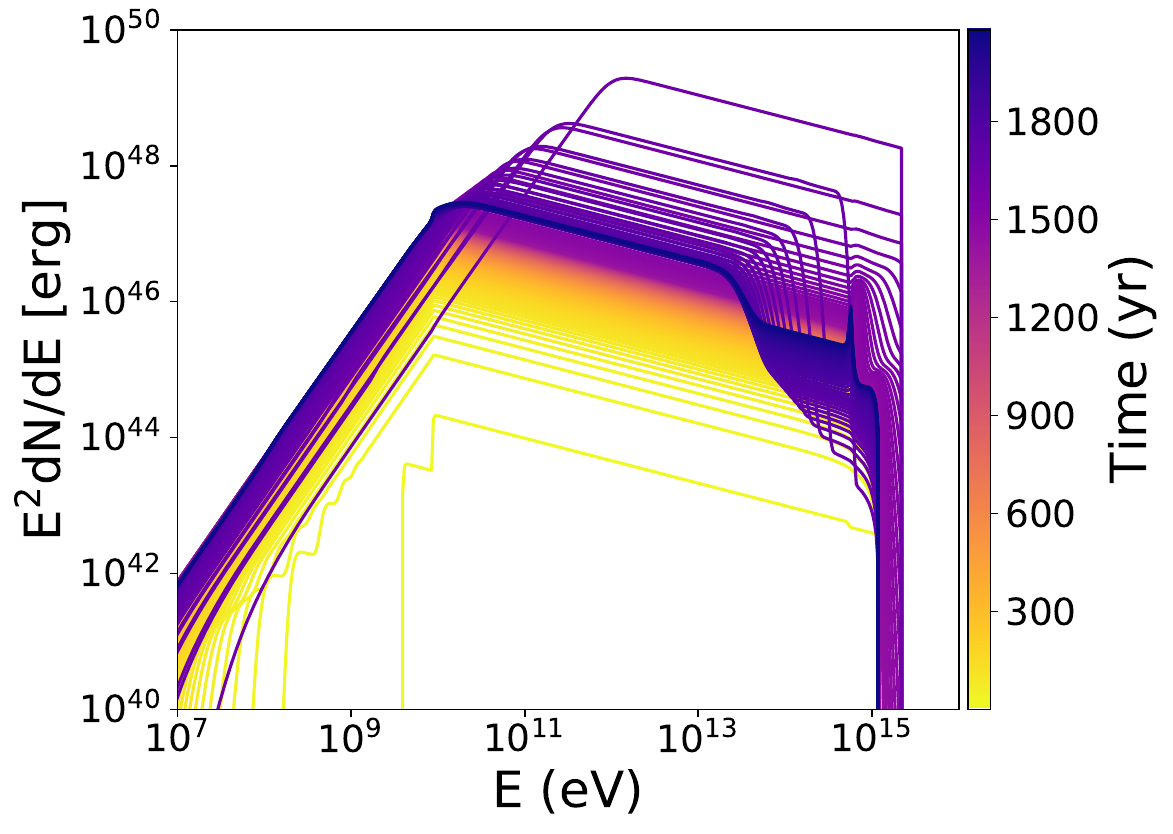}
        \includegraphics[width=0.49\textwidth]{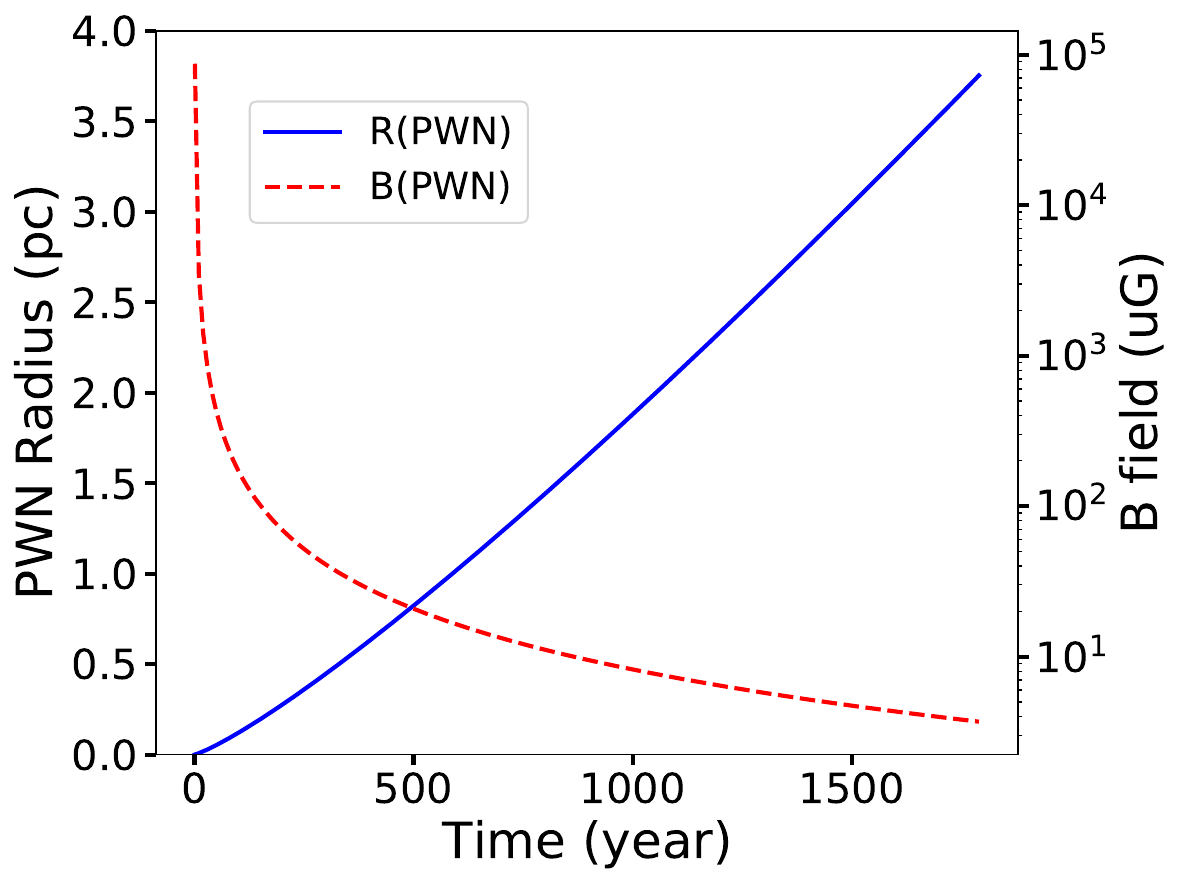}
        \includegraphics[width=0.49\textwidth]{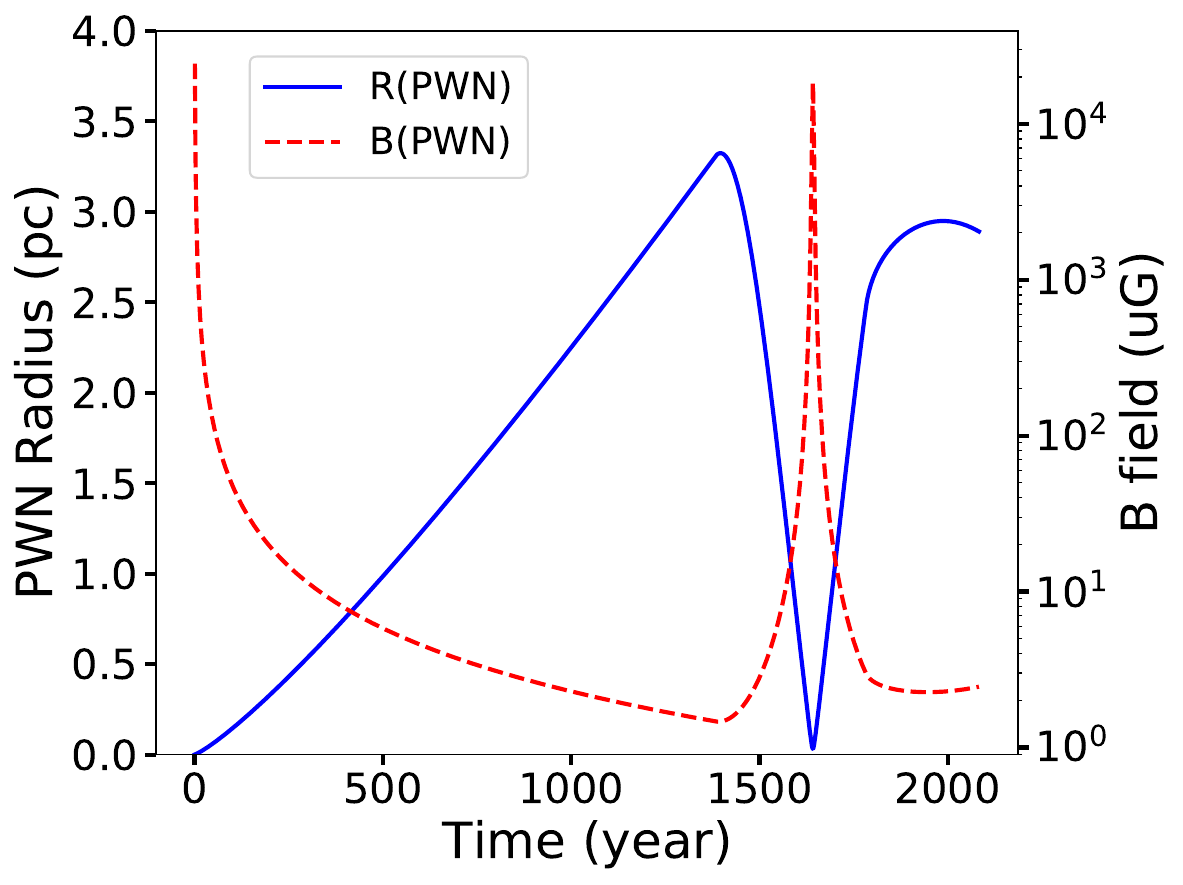}
    \caption{Top panel: Electron evolution for Case B (left) and Case C (right) of Figure \ref{fig:sed_evolution} and Table \ref{tab:evolutionary_table}. Bottom panel: $R_{pwn}$ and magnetic field evolution of Case B (left) and Case C (right).} 
    \label{fig:evolutionary_size_bfield}
\end{figure*}

(2) $d = 7.5$ kpc case: Despite the larger distance, we still find it constraining to fit the SED data and PWN size simultaneously. Below we present two different cases: (1) a young PWN in the free-expansion phase (Case B) and (2) a re-expanding PWN after SNR crush (Case C). Both cases have the same number of free parameters. In both cases, we need to evolve the PWN size to match $r\sim4$ pc while keeping the number of injected leptons to the level required for fitting the radio, X-ray and gamma-ray SED data. The 3rd column in Table \ref{tab:evolutionary_table} and Figure \ref{fig:sed_evolution} (upper right panel) show the model parameters and SED plot, respectively, for Case B. This case assumes that the PWN has been expanding over 1.8 kyr, and it is similar to the {\tt GAMERA} model for $d = 7.5$ kpc. As Figure \ref{fig:evolutionary_size_bfield} (upper panels) shows, the current B-field is 3.7 $\mu$G while the PWN expanded to $r=3.8$ pc. Note that the power-law spectral index is softer below than above $E_{\rm break}$ in Case B, which is highly unusual, but has been seen in a few other cases \citep{Hattori2020, Temim2015}. The spectral break at the highest radio flux point in Model B is caused by the minimum energy of the particle injected at the termination shock. Below this energy, all of the particles were injected earlier and cooled to this regime. Above this energy, there are a mix of particles injected earlier which have cooled, and freshly injected particles that haven't had time to cool yet. The energy spectrum of these two particles populations are different, which leads to a change in slope at the energy separating the two. As can be seen in Figure \ref{fig:sed_evolution}, Case B fits the X-ray data well, while a clear trend away from the model can be seen in the residuals plot in the radio band. Since the emission in the radio band is dominated by low energy particles, this discrepancy between data and model may be due to our simplified assumption that $E_{\rm min}$ stays constant over time. In contrast to Case B, Case C represents the SNR crush scenario, as was suggested by \citet{Kothes2006}. We evolved the PWN through the reverse shock crush into re-expansion over $t_{\rm age} = 2.1$ kyr, as indicated by the PWN radius plot over time in Figure \ref{fig:evolutionary_size_bfield} (lower panels). The current B-field and PWN radius are $2.5~\mu$G and 2.9 pc, respectively. As seen in Figure \ref{fig:sed_evolution}, the model fits the data well. While slightly overshooting the \veritas\ upper limit, factoring in the $\sim 20\%$ uncertainty in the upper-limit of the TeV photon density at the decorrelation energy due to the unknown photon index in this band, the model is consistent with the non-detection by \veritas. Given how close the minimum and break particle energies are, the model favors a single power-law injection spectrum with $p\sim2.3$ (i.e., standard Fermi).

Reproducing the gamma-ray emission of many PWNe often requires background photon fields in addition to the CMB -- either from surrounding dust
\citep[e.g., G21.5-0.9; ][]{Hattori2020} and/or nearby stars \citep[e.g., HESS J1640-465 and Kes 75; ][]{moaz2022, Straal2023}. Since the gamma-ray emission from the Boomerang has not been detected, the presence of additional background photon fields cannot be directly constrained by the modeling above. Therefore, to determine their possible effect on the parameters derived above, we modeled the SED and dynamical properties of this source assuming an additional photon field with a temperature T = 30 K and energy density 5x that of the CMB, typical values for warm dust in these systems \citep{Cox1986, Torres2013}. As shown in Table \ref{tab:evolutionary_table}, for Case B' we can reproduce the properties of the Boomerang for a set of model parameters similar to that of Case B (where the only background photon field is the CMB). However, in our parameter exploration we were not able to reproduce the observed properties of the Boomerang, assuming this additional background photon field, for a set of parameters similar to Case C in Table \ref{tab:evolutionary_table}.  This does not necessarily exclude Case C as a reasonable description for the Boomerang, since there are many regions in the Galaxy where the energy density of dust emission is less than that of the CMB \citep[e.g., G21.5-0.9; ][]{Strong2000}.

Overall, the PWN evolution model does not reproduce both the SED data and PWN size with reasonable parameters if we assume $d = 0.8$ kpc. In all of the SED models presented in this section, we found $B \sim 2\rm{-}4~\mu$G. 
Alternatively, we found that an extremely magnetized PWN with $\eta_B \sim 1$ can fit the SED data well. More specifically, when we adopt $\eta_B = 0.99$, only 1\% of the pulsar's rotational energy is allocated to particle injection and the current B-field is $\sim 100~\mu$G. The smaller number of injected leptons and higher B-field cancel with each other to fit the synchrotron SED in the radio and X-ray bands. However, such a high magnetization parameter is unusual compared to those of six other PWNe ($\eta_B = 7\times10^{-4}\rm{-}0.02$) including the Crab nebula \citep{Martin2014A}. 
For the case of $d=7.5$ kpc, we consider Case C more plausible since Case B and Case B' suggest that the high-energy particle index is greater than the low energy particle index. Furthermore, in Case C the PWN interacts with the SNR reverse shock, which could explain the offset between the radio and X-ray peak emission. No explanation for this offset can be inferred from Case B or Case B'.

\section{Discussion} 
\label{sec:Discussion}

We discuss constraining the properties of the Boomerang PWN and its recent evolution based on the X-ray and multi-wavelength observations. In the previous sections, we examined the hypothesis proposed from the radio observations \citep{Kothes2006} that the Boomerang is a highly magnetized PWN ($B = 2.6$ mG) crushed by an SNR reverse shock 3,900 yr ago, which was made under the assumption that the radio break in Boomerang's spectrum is due to synchrotron cooling. However, the radio break is not necessarily caused by synchrotron cooling, which may explain the discrepancy between the results in this paper and those from \citet{Kothes2006}. 

Below, we present some implications for this composite SNR--PWN system, especially in the head region, and suggest future observations to further elucidate the origin of the head--tail morphology and UHE emission.

\subsection{Constraining PWN magnetic field} 
\label{sec:pwn_size} 

As shown in \S\ref{sec:SED}, our SED study strongly suggests that the current B-field should be $B\simlt 3 \mu$G, otherwise the observed synchrotron fluxes will be significantly over-predicted. Combined with the {\tt GAMERA} SED model results in \S\ref{sec:GAMERA}, below we consider the energy-dependent X-ray size measurements from the \nustar\ observation and constrain the PWN B-field. The smaller X-ray size (e.g., $r=20$\asec\ in 10--20 keV) compared to the radio size ($r=100$\asec) is often observed for other PWNe \citep{Coerver2019}. The radio nebula size usually reflects the PWN size determined by the particle flow evolution over the pulsar's age. On the contrary, the smaller X-ray size is determined by synchrotron cooling time. 
The synchrotron cooling time for electrons emitting X-rays of energy $E$ [keV] is $\tau_{\rm syn} = 1.2 B_{\rm mG}^{-3/2} E_{\rm keV}^{-1/2}$ yr \citep{Reynolds2018}. Since the synchrotron cooling time depends on electron energy, we expect X-ray PWN size to shrink with photon energy. The synchrotron burn-off effect is evident, as we measured different X-ray sizes in two energy bands -- 33\asec\  (3--10 keV) and 20\asec\ (10--20 keV). Assuming a constant flow velocity over the X-ray synchrotron cooling time, the ratio of the PWN radii between the hard and soft energy bands should be approximately equal to the ratio between the synchrotron cooling times for the respective mid-band energies. Subsequently, adopting the mid energy in each band (6.5 and 15 keV), we estimate that the PWN size ratio between the two energy bands should be 1.5 if the flow velocity is constant, which is close to the observed ratio of 1.7. Hence, we consider the constant flow velocity hypothesis as viable. 

Assuming a source distance of 7.5 kpc (as suggested by the pulsar's DM measurement and supported by our SED model fitting), the X-ray angular sizes in the 3--10 and 10--20 keV bands correspond to PWN radii of $r\sim0.7$ pc and $\sim0.4$ pc, respectively. As expected, these radii are significantly smaller than the PWN radius assumed for the SED modelling ($\sim$4 pc at $d=7.5$ kpc), which represents the size of the PWN in the radio band.
For a given B-field, we can derive an upper limit of the PWN size assuming that leptons moved outward at the speed of light during their lifetime. Using the synchrotron lifetime for the highest energy X-ray emission ($E_\gamma = 20$ keV), the hard X-ray PWN size of $r=0.4$ pc sets an upper limit of $B <  0.35$ mG, which is lower than the B-field value suggested by \citet{Kothes2006}. We can further constrain B-field with a more realistic estimate of the PWN flow velocity. For example, multi-epoch \chandra\ X-ray observations of Kes-75 measured an expansion of the PWN at $V_{\rm PWN} \sim 1,000$ km\,s$^{-1}$ or $3.3\times10^{-3}c$ \citep{Reynolds2018}. The Crab nebula is known to expand at a velocity of $V_{\rm PWN} \sim 1,500$ km\,s$^{-1}$ \citep{Bietenholz1991} and its radius increases almost linearly with time -- $R(t) \propto t^{1.264}$ \citep{Bietenholz2015}. 
If we adopt these flow velocities, the hard X-ray size of the Boomerang PWN in the 10--20 keV band suggests $B\sim 7\rm{-}10~\mu$G. 
We also estimated a range of the flow velocity using the output PWN size evolution data (i.e., $R(t)$) provided by the dynamical PWN model (\S\ref{sec:DynamicSED}). 
We found $v_{\rm PWN} \sim 2,400$ km/s and $\sim400$ km/s over the last 60 years of expansion for Case B and C, respectively, and their PWN radius evolution is shown in Figure \ref{fig:evolutionary_size_bfield}. These PWN velocities yield $B \sim 6\rm{-}14~\mu$G, which is slightly higher than the B-field suggested by the dynamical model fitting in \S\ref{sec:DynamicSED}.

Our estimates for the PWN B-field, based on the PWN's X-ray spatial properties and SED fitting, are not only significantly below what was suggested by \cite{Kothes2006} but also B = 140~$\mu$G, which was suggested by \cite{Liang2022} based on modelling of the diffusion of relativistic electrons in the PWN. Our disagreement with the B-field derived from the analysis done by \cite{Liang2022} is expected given our different assumptions and models. While \cite{Liang2022} incorporated UHE tail emission not necessarily related to the PWN itself within their SED modelling, this paper focuses solely on emission within the general bounds of the PWN region. Furthermore, unlike the dynamical model used in this paper, the PWN model used in \cite{Liang2022} does not include any interaction with the SNR, and the temporal evolution consists only of changes resulting from diffusion and radiative losses.

\subsection{PWN evolution} 

The dynamical SED model fitting \S\ref{sec:DynamicSED} suggests that the Boomerang PWN is currently re-expanding to $r\sim3$~pc after being crushed by a SNR reverse shock $\sim1000$ yr ago, much shorter than the pulsar's $\sim10$ kyr characteristic age. As can be seen in the Case C scenario of the bottom panels of Figure \ref{fig:evolutionary_size_bfield}, we note that the PWN is undergoing a small second compression due to an overshoot in its re-expansion, which resulted in a negative pressure differential between the PWN's interior and surrounding material.
The Boomerang PWN is powered by a population of fresh electrons injected over the last $\sim1000$ yr. After the PWN compression amplified B-field to $B \simgt 10$ mG, its B-field decreased, as of present day, to $B\sim 3~\mu$G as a result of the nebula expansion. It is interesting to note the consequence of this temporally varying B-field on the spatial variations of the PWN's spectrum. Particles are generally older further away from their source pulsar \citep{Temim2015}. Subsequently, if the B-field was higher in the past as suggested by the dynamical model, leptons further away from \psr\ would generally have experienced more significant synchrotron losses as compared to those found closer to \psr. Therefore, this decrease in B-field over time may explain the significant spectral softening in the X-ray spectrum shown to occur with increasing distance from the pulsar by \cite{Ge2021}.

Hydrodynamic (HD) simulation of SNR-PWN interaction exhibited head and tail-like features in the particle density maps \citep[Figure 2 in ][]{Kolb2017}. According to this simulation study, there are two factors that determine the composite SNR--PWN morphology -- (1) density gradient and (2) pulsar proper motion. 

As suggested by \citet{Kothes2006}, the density gradient around the head region likely  caused their highly asymmetric morphology and Boomerang-like PWN shape in the radio band. While the X-ray bright PWN ($r \sim 0.4\rm{-}0.7$ pc) should be powered by a population of young electrons injected over the last $\sim 1000$ yr, the relic electrons injected in prior to the SNR crush should still be producing synchrotron radiation in the radio and X-ray band. We would associate the relic PWN radiation with the head region, while we attribute the lack of TeV emission in the PWN region to a small number of electrons injected by the ``refreshed'' PWN.  Since the total number of relic electrons injected over a few thousand years is much higher, the head region should produce higher TeV emission via ICS than in the PWN. The reported TeV detection in the head region by MAGIC may indicate such ICS emission. The UHE emission in the tail region cannot be caused by relic electrons which should have cooled quickly to GeV--TeV energies. Instead, particle re-acceleration during the SNR-PWN interaction via PWN compression, as proposed by \citet{Ohira2018}, may be responsible for producing the UHE emission in the tail. A further study in the entire region, with new VERITAS observation data and more extensive SED study, will be presented in our future paper. 

Another key parameter is the proper motion of the pulsar. Although there is yet no direct measurement of the proper motion, the radio and X-ray morphology of the PWN suggest that the pulsar is moving in the north--west direction. It was proposed that the Boomerang-like radio morphology was caused by the PWN colliding into the high density ISM region in the north--west boundary of SNR~G106.3$+$2.7. In this scenario, the head is more extended because the relic electrons from the PWN diffuse into a lower density region. In addition, the X-ray torus--jet morphology data  was fit by a PWN emission model \citep{Ng2004} also supporting the proper motion in the north--west direction. Although a  future \chandra\ observation over $\sim 20$ year baseline may be able to detect the proper motion, it may be difficult if the source distance is indeed $\sim8$ kpc or greater as suggested by our SED and morphology studies presented in this paper.

\section{Conclusions} 
\label{sec:Conclusions} 

We summarize our findings from our multi-wavelength investigation of the Boomerang PWN and its surrounding region. 

\begin{itemize} 

\item[(1)] We detected a 51.67 ms pulsation with 3--20 keV \nustar\ data and it allowed us to separate the on-pulse component from the PWN emission. With \nustar\, we detected hard X-ray emission from the Boomerang PWN up to 20 keV and found that its size decreases from $r=33\pm2\asec$ (3--10 keV) to $20\pm2\asec$ (10--20 keV). 

\item[(2)] Our analysis of the 2002 \chandra\ data, after excising the pulsar emission, yielded $N_{\rm H} = 8.9^{+1.5}_{-1.4}\times10^{21}$ cm$^{-2}$ and the PWN photon index of $\Gamma = 1.52^{+0.13}_{-0.12}$. 
The hydrogen column density is consistent with the value derived from the pulsar's DM measurement of $(4.3\rm{-}8.8)\times10^{21}$ cm$^{-2}$. 
Joint \chandra\ and \nustar\ spectral analysis of the PWN measured no X-ray flux variability and the 0.5--20 keV spectra are consistent with a single power-law model with $\Gamma = 1.52 \pm 0.06$. 

\item[(3)] Our analysis of the most updated \fermi\ and \veritas\ data yielded no detection of the Boomerang PWN. We set an upper limit on the gamma-ray flux above 50 GeV to $F \simlt 10^{-12}$ \fluxcgs.

\item[(4)] Among the previously suggested source distances ranging from 0.8 kpc to 7.5 kpc \citep{Kothes2001, Abdo2009b, He2013}, we found that $d \sim 8$ kpc provides the most plausible solution to fitting the SED data. The widely-used source  distance of 0.8 kpc leads to too low radiation efficiencies that cannot be produced by any of our SED models with reasonable parameters. We note, however, that all SED models used in this paper simplify the complex nature of Boomerang. While the offset between the radio and X-ray peak emission may suggest a multi-zone system, modelling was done under the assumption that Boomerang is a single-zone system. Implementing an multi-zone model may boost our understanding of the emission offset between the radio, X-ray, and TeV peak emission. The dynamical SED model used in this paper also assumes that Boomerang is a spherically symmetric PWN, an obvious simplification of the actual morphology. As can be seen in Figure \ref{fig:sed_evolution}, the dynamical SED fits predict strong emission at slightly higher radio frequencies and/or X-ray energies than what has been observed so far. Future analysis of observations in these energy bands could further test the validity of these fits.

\item[(5)] The radiation efficiencies (i.e. $F(4\pi R^{2})/\dot{E}$ ratios) of Boomerang are $3\times10^{-6}(\frac{R}{8~kpc})^{2}$, $6\times10^{-4}(\frac{R}{8~kpc})^{2}$ and $4\times10^{-4}(\frac{R}{8~kpc})^{2}$ in the radio, X-ray and TeV bands, respectively. Assuming a distance of 8 kpc, these values are now more consistent with those of other PWNe with similar spin-down powers \citep{Kargaltsev2013}.  

\item[(6)] Our SED modelling requires that the current PWN B-field should be low ($B\sim 2\rm{-}4~\mu$G). Our X-ray PWN size measurements suggest a slightly higher value ($B\sim 6\rm{-}14~\mu$G). We ruled out the high B-field value ($B = 2.6$ mG) suggested by the radio observations \citep{Kothes2006}, as well as the B-field value (B = 140~$\mu$G) suggested by \cite{Liang2022}.

\item[(7)] The Boomerang PWN should currently be in its re-expansion phase after having been crushed by the SNR reverse shock. The relic electrons injected earlier before the SNR crush should have produced synchrotron and ICS radiation elsewhere, e.g. in the head region. We attribute the radio break at 4--5 GHz to the minimum energy of the injected electrons. While the origin of this break in the electron spectrum is still unknown, we are confident that the radio break is not caused by synchrotron cooling as was hypothesized by \citet{Kothes2006}. 

\item[(8)] The origin of the head--tail morphology could be related to the PWN propagating in an inhomogeneous density region and interacting with the SNR, as suggested by HD simulations. Following \citet{Kothes2006}, we also suspect that the PWN is currently propagating in the northwest direction and expanding in high-density ISM. 
In order to understand the origin of VHE and UHE gamma rays and their connection to the SNR-PWN system, it would be important to resolve the spatial distribution in the TeV emission in the head and tail regions as well as determine the distance and proper motion of the \psr.   

\end{itemize}

\begin{acknowledgments}
Support for this work was provided by NASA through NuSTAR Cycle
6 Guest Observer Program grant NNH19ZDA001N. VERITAS is supported by grants from the U.S. Department of Energy Office of Science, the U.S. National Science Foundation and the Smithsonian Institution, by NSERC in Canada, and by the Helmholtz Association in Germany. This research used resources provided by the Open Science Grid, which is supported by the National Science Foundation and the U.S. Department of Energy's Office of Science, and resources of the National Energy Research Scientific Computing Center (NERSC), a U.S. Department of Energy Office of Science User Facility operated under Contract No. DE-AC02-05CH11231. We acknowledge the excellent work of the technical support staff at the Fred Lawrence Whipple Observatory and at the collaborating institutions in the construction and operation of the instrument.
\end{acknowledgments}

\facilities{NuSTAR, Chandra, Fermi-LAT, VERITAS}

\software{HEAsoft (v6.28), NuSTARDAS (v2.0.0), XSPEC (v12.11.1; \cite{Arnaud1996}), CIAO (v4.13), Naima \citep{Zabalza2015}}, GAMERA \citep{Hahn2016} 


\bibliography{citations}{}
\bibliographystyle{aasjournal}



\end{document}